\documentclass[fleqn,usenatbib]{mnras}
\usepackage{amsmath}    
\usepackage{newtxtext,newtxmath}
\usepackage[T1]{fontenc}
\usepackage{ae,aecompl}
\usepackage{graphicx}	
\usepackage{geometry}
\usepackage{float}
\usepackage{caption}
\usepackage{subcaption}
\usepackage{dblfloatfix}
\usepackage{url}
\usepackage{epstopdf}
\newcommand{\hinvMsun}{h^{-1}{\rm M}_\odot}
\newcommand{\hinvMpc}{h^{-1}{\rm Mpc}}
\newcommand{\Vmax}{$V_{\rm max}$}
\newcommand{\Vpeak}{$V_{\rm peak}$}
\newcommand{\ahalf}{$a_{\rm M/2}$}
\newcommand{\Rvir}{$R_{\rm vir}$}
\newcommand{\Vp}{V_{\rm peak}}
\newcommand{\Mdot}{\dot M_{\rm h}}
\newcommand{\Mh}{M_{\rm h}}

\newcommand{\Mstar}{M_*}
\newcommand{\Msun}{{\rm M}_\odot}

\title{Galaxy assembly bias of central galaxies in the Illustris simulation}
\author[Xu \& Zheng]{
Xiaoju Xu\thanks{E-mail: xiaoju.xu@utah.edu}
and Zheng Zheng\thanks{E-mail: zhengzheng@astro.utah.edu}\\
\\
Department of Physics and Astronomy, University of Utah, 115 South
1400 East,
Salt Lake City, UT 84112, USA
}
\date{Accepted XXX. Received YYY; in original form ZZZ}

\pubyear{2018}

\begin{document}
\label{firstpage}
\pagerange{\pageref{firstpage}--\pageref{lastpage}}
\maketitle

\begin{abstract}
Galaxy assembly bias, the correlation between galaxy properties and halo properties at fixed halo mass, could be an important ingredient in halo-based modelling of galaxy clustering. We investigate the central galaxy assembly bias by studying the relation between various galaxy and halo properties in the Illustris hydrodynamic galaxy formation simulation. Galaxy stellar mass $\Mstar$ is found to have a tighter correlation with peak maximum halo circular velocity $\Vp$ than with halo mass $\Mh$. Once the correlation with $\Vp$ is accounted for, $\Mstar$ has nearly no dependence on any other halo assembly variables. The correlations between galaxy properties related to star formation history and halo assembly properties also show a cleaner form as a function of $\Vp$ than as a function of $\Mh$, with the main correlation being with halo formation time and to a less extent halo concentration. Based on the galaxy-halo relation, we present a simple model to relate the bias factors of a central galaxy sample and the corresponding halo sample, both selected based on assembly-related properties. It is found that they are connected by the correlation coefficient of the galaxy and halo properties used to define the two samples, which provides a reasonable description for the samples in the simulation and suggests a simple prescription to incorporate galaxy assembly bias into the halo model. By applying the model to the local galaxy clustering measurements in Lin et al. (2016), we infer that the correlation between star formation history or specific star formation rate and halo formation time is consistent with being weak.
\end{abstract}


\section{Introduction}
\label{sec:intro}

It has been well established that galaxies form in dark matter haloes \citep{White1978}. As the first step to study galaxy formation and clustering, halo formation and clustering, which is dominated by gravity, have been extensively studied with analytic models \citep[e.g.][]{Press74,Bardden86,Mo96,Sheth99} and cosmological $N$-body simulations \citep[e.g.][]{Springel05,Prada12}. It has been found that halo clustering depends not only on halo mass but also on halo assembly history and environment \citep[e.g.][]{Gao05,Gao07,Paranjape18,Xu18,Han19,Ramakrishnan19}, or halo definition \citep[e.g.][]{Villarreal17, Mansfield19}. This is called halo assembly bias, whose nature is still under investigation \citep[e.g.][]{Dalal08,Castorina13,Shi18}.

If galaxy properties are affected by halo formation and assembly history, halo assembly bias would translate to galaxy assembly bias. Operationally, galaxy assembly bias can be defined as that at fixed halo mass, the statistical galaxy content shows dependence on other halo variables or galaxy properties show correlations with halo assembly history (which would cause a certain degree of halo assembly bias to be inherited by galaxies and to show up in galaxy clustering). The widely adopted halo model \citep[e.g.][]{Cooray02} of interpreting galaxy clustering, such as the halo occupation distribution  \citep[e.g.,][]{Berlind02,Zheng05} or conditional luminosity function \citep[e.g.,][]{Yang03}, makes the implicit assumption of no galaxy assembly bias. Such methods have been successfully applied to galaxy clustering \citep[e.g.][]{Zehavi05,Zheng07,HXu18}. However, if assembly bias is significant, neglecting it in the model would lead to incorrect inference of galaxy-halo connections and introduce possible systematics in cosmological constraints (e.g. \citealt{Zentner14,Zentner16}; but see also \citealt{McEwen16,McCarthy18}). Conversely, observationally inferred galaxy assembly bias would help understand galaxy formation.

The existence and strength of galaxy assembly bias are still a matter far from settled, either in theory or in observation. Galaxy assembly bias has been investigated in hydrodynamic \citep[e.g.][]{Berlind03,Mehta14,Chaves-Montero16,Artale18,Martizzi19,Bose19} or semi-analytic galaxy formation models \citep[e.g.][]{Croton07,Contreras18,Zehavi18,Zehavi19}, focusing on the effect on galaxy occupation function and galaxy clustering. The results seem to depend on the implementation details of star formation and feedback. Studying galaxy assembly bias from observation has the difficulty of determining halo mass, and the results are not conclusive \citep[e.g.][]{Yang06,Berlind06,Lin16,Zu17,Guo17}. Given the potential importance of galaxy assembly bias in modelling galaxy clustering, in this paper we study the correlation between various central galaxy properties and halo properties in the Illustris hydrodynamic simulation \citep{Vogelsberger2014a} at the halo level, aiming at providing useful insights in describing galaxy assembly bias.

The structure of the paper is as follows. In section ~\ref{sec:sim}, we introduce the simulation and the galaxy and halo catalogues. Then in section~\ref{sec:result}, we investigate the relation between galaxy and halo properties,  with primary galaxy-halo properties in section ~\ref{sec:mstardep} and general galaxy-halo properties in section ~\ref{sec:ghrelation}. In section ~\ref{sec:gab}, we present a simple model to connect galaxy assembly bias with halo assembly bias. Finally, we summarise and discuss the results in section ~\ref{sec:summary}.

\section{Simulation and galaxy-halo catalogue}
\label{sec:sim}

In this work, we use galaxies and haloes from the state-of-the-art hydrodynamic galaxy formation simulation Illustris\footnote{http://www.illustris-project.org} \citep{Vogelsberger2014a,Nelson15} to study galaxy assembly bias, which is able to produce different type of galaxies seen in observation \citep{Vogelsberger2014b,Genel14}. In particular, we use the Illustris-2 simulation, which has a box size of 75$h^{-1}$Mpc on a side, and contains $910^3$ dark matter particles of mass $5\times 10^7\Msun$ and the same number of baryon particles of mass $1\times 10^7\Msun$. The mass resolution is sufficient for our purpose of studying (central) galaxies in haloes of more massive than a few times $10^{10}\hinvMsun$. The simulation adopts a spatially-flat cosmology with the following parameters: $\Omega_{\rm m}=0.27$, $\Omega_{\rm b}=0.0456$, $h=0.70$, $n_{\rm s}=0.963$, and $\sigma_8=0.809$. 

The haloes in the Illustris database are identified with the friends-of-friends (FoF) algorithm. As this algorithm is notorious for
having the probability of bridging two separate halos into one halo, we apply the phase-space halo finder Rockstar \citep{Behroozi13a} to identify haloes, with dark matter particles extracted from 30 snapshots of the simulation (from z=3.94 to z=0). We then build the merger tree using the consistent-tree algorithm \citep{Behroozi13b}.

We focus our study on the relation between central galaxies and dark matter haloes and the assembly effect at $z=0$. Galaxies in the Illustris-2 simulation are assigned to the Rockstar haloes. For each galaxy, if its distance $d_{gh}$ to the centre of a halo is smaller than the virial radius \Rvir\ of the halo, it is assigned to this halo. In rare cases, a galaxy may be assigned to more than one haloes, as halos are not spherical. We then put the galaxy into the halo that corresponds to the lowest ratio of $d_{gh}/$\Rvir. With galaxies assigned to haloes, for each halo we define the most massive galaxy inside 0.2\Rvir\ from the halo centre to be the central galaxy. If there is no galaxy inside 0.2\Rvir, we simply choose the most massive galaxy inside \Rvir\ as the central galaxy.

With the catalogue of central galaxies and the associated host haloes, we study the relation between galaxy properties and halo properties and the assembly effects. The halo properties we focus on are: 
\begin{itemize}
    \item[(1)] $\Mh$, halo mass enclosed in a volume with mean density of 200 times the background density of the universe; 
    \item[(2)] \Vpeak, peak maximum circular velocity of the halo over its accretion history; 
    \item[(3)] $c$, halo concentration parameter, defined as the ratio of halo virial radius to scale radius; 
    \item[(4)] \ahalf, cosmic scale factor when the halo obtains half of its current ($z=0$) total mass; 
    \item[(5)] $\Mdot$, halo mass accretion rate near $z=0$ (averaged between z=0 and z=0.197, about 2.4 Gyr, one dynamical time), in units of $\hinvMsun\, {\rm yr}^{-1}$; 
    \item[(6)] $\Mdot/\Mh$, specific halo accretion rate, in units of ${\rm Gyr}^{-1}$. 
\end{itemize}
The central galaxy properties we consider include: 
\begin{itemize}
    \item[(1)] $\Mstar$, stellar mass (sum of masses of star particles within twice the stellar half mass radius); 
    \item[(2)] SFR, star formation rate within twice the stellar half mass radius; 
    \item[(3)] sSFR, specific star formation rate, the ratio of SFR to $\Mstar$; 
    \item[(4)] $g-r$, galaxy colour defined by the $g$-band and $r$-band luminosity.
\end{itemize}

\section{Results}
\label{sec:result}

We aim at presenting the relation between central galaxies and haloes to learn about the correlation between halo formation and assembly and galaxy properties. In section~\ref{sec:mstardep} We first study how galaxy stellar mass depends on the primary halo properties ($\Mh$ and \Vpeak). Then we investigate how various halo and galaxy properties are correlated in section~\ref{sec:ghrelation}. Finally, in section~\ref{sec:gab} we use a simplified model to describe the connection between galaxy and halo assembly bias factor.

\subsection{Relationship between stellar mass and halo properties}
\label{sec:mstardep}

\begin{figure*}
	\centering
	\begin{subfigure}[h]{0.48\textwidth}
		\includegraphics[width=\textwidth]{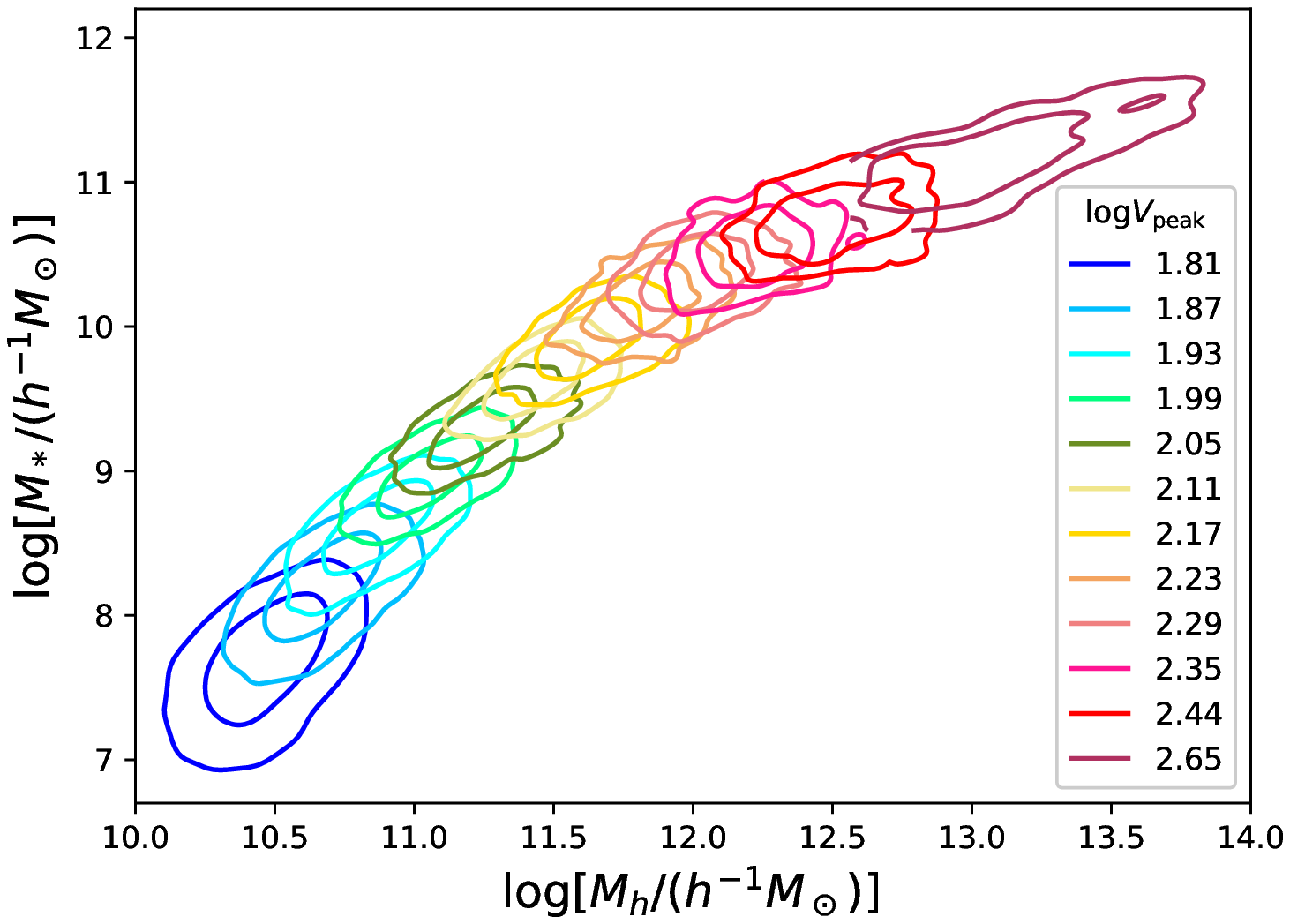}
	\end{subfigure}
	\hfill
	\begin{subfigure}[h]{0.48\textwidth}
        \includegraphics[width=\textwidth]{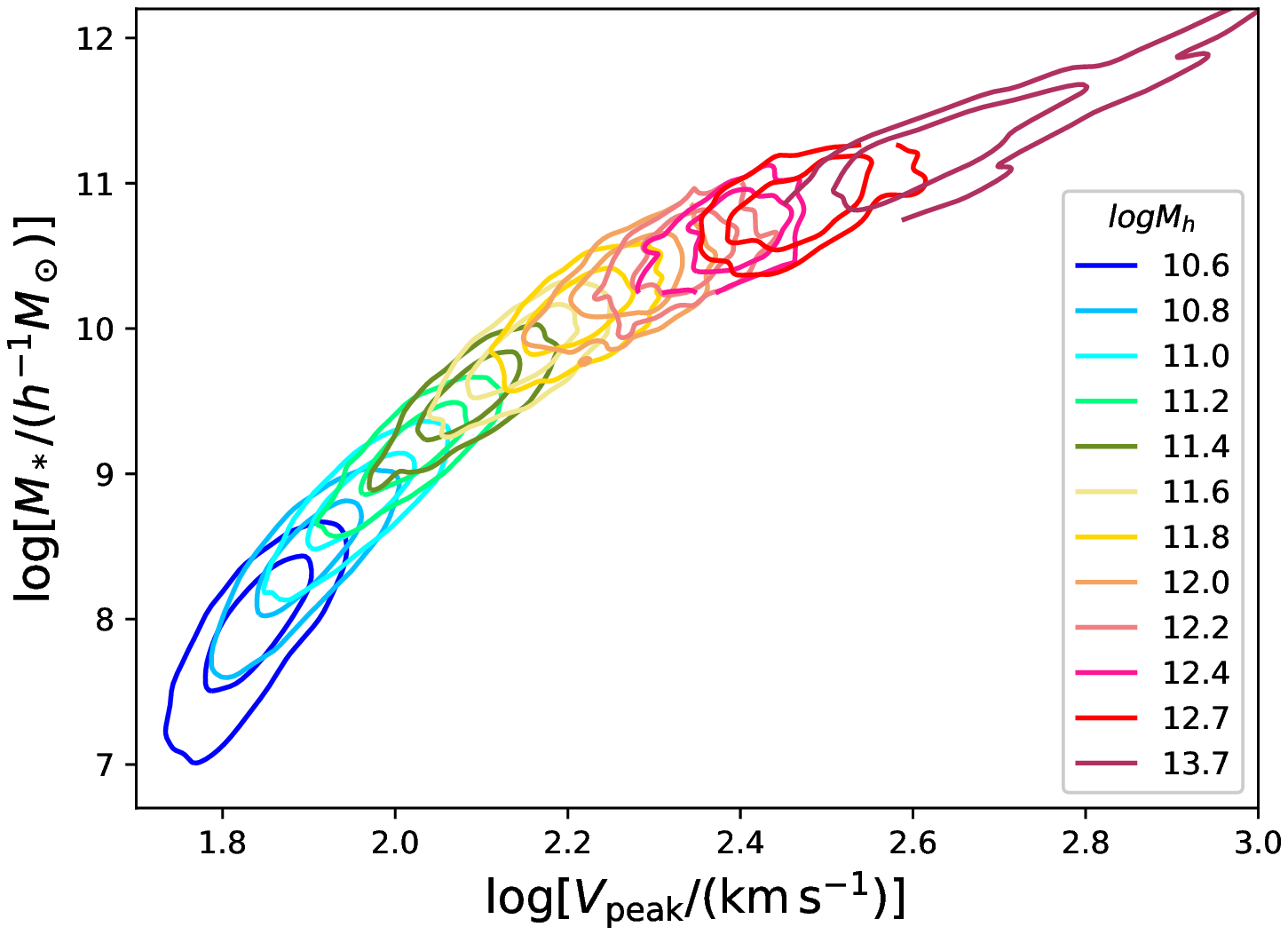}
	\end{subfigure}
	\hfill
	\begin{subfigure}[h]{0.48\textwidth}
		\includegraphics[width=\textwidth]{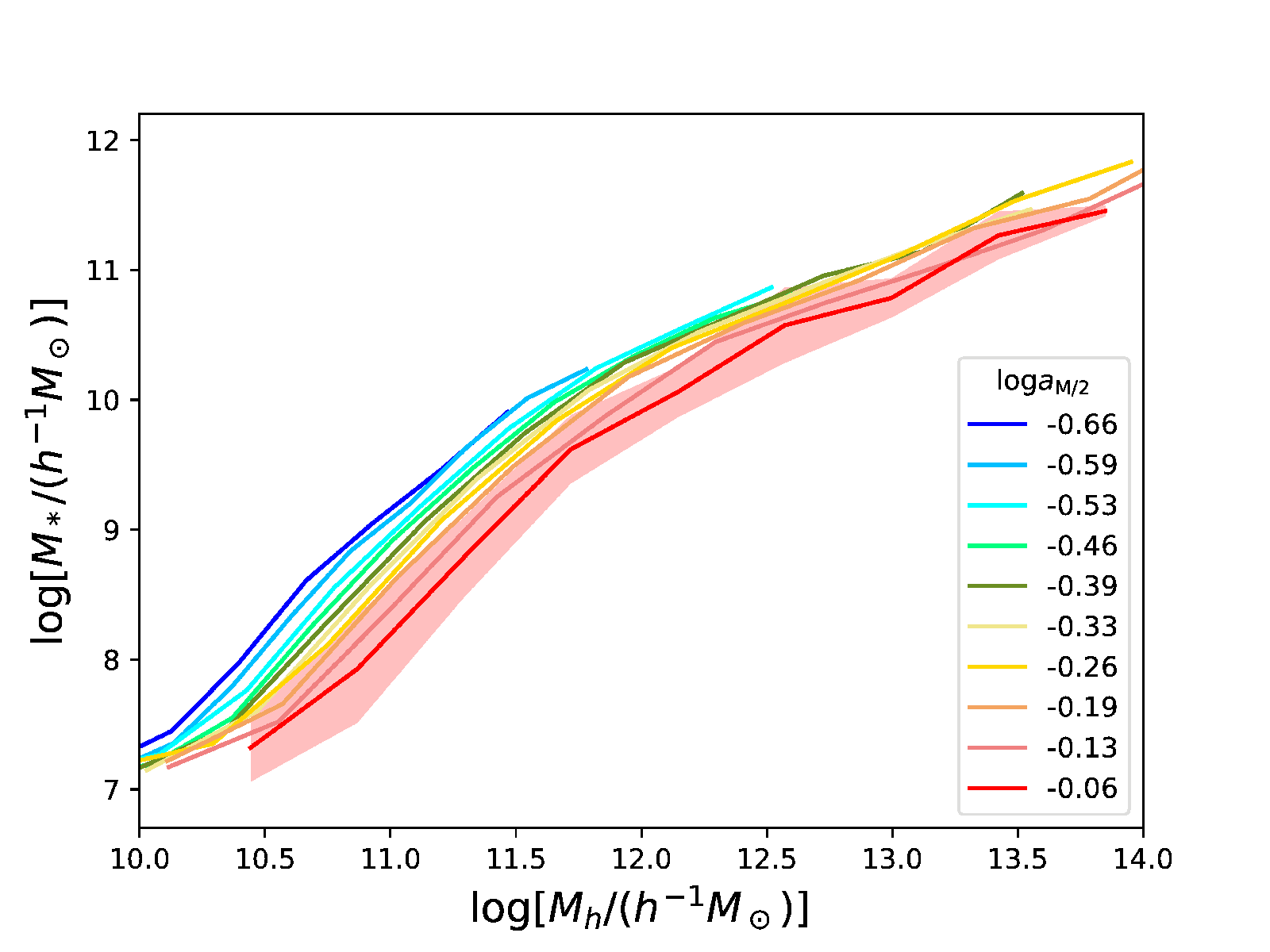}
	\end{subfigure}
	\hfill
	\begin{subfigure}[h]{0.48\textwidth}
		\includegraphics[width=\textwidth]{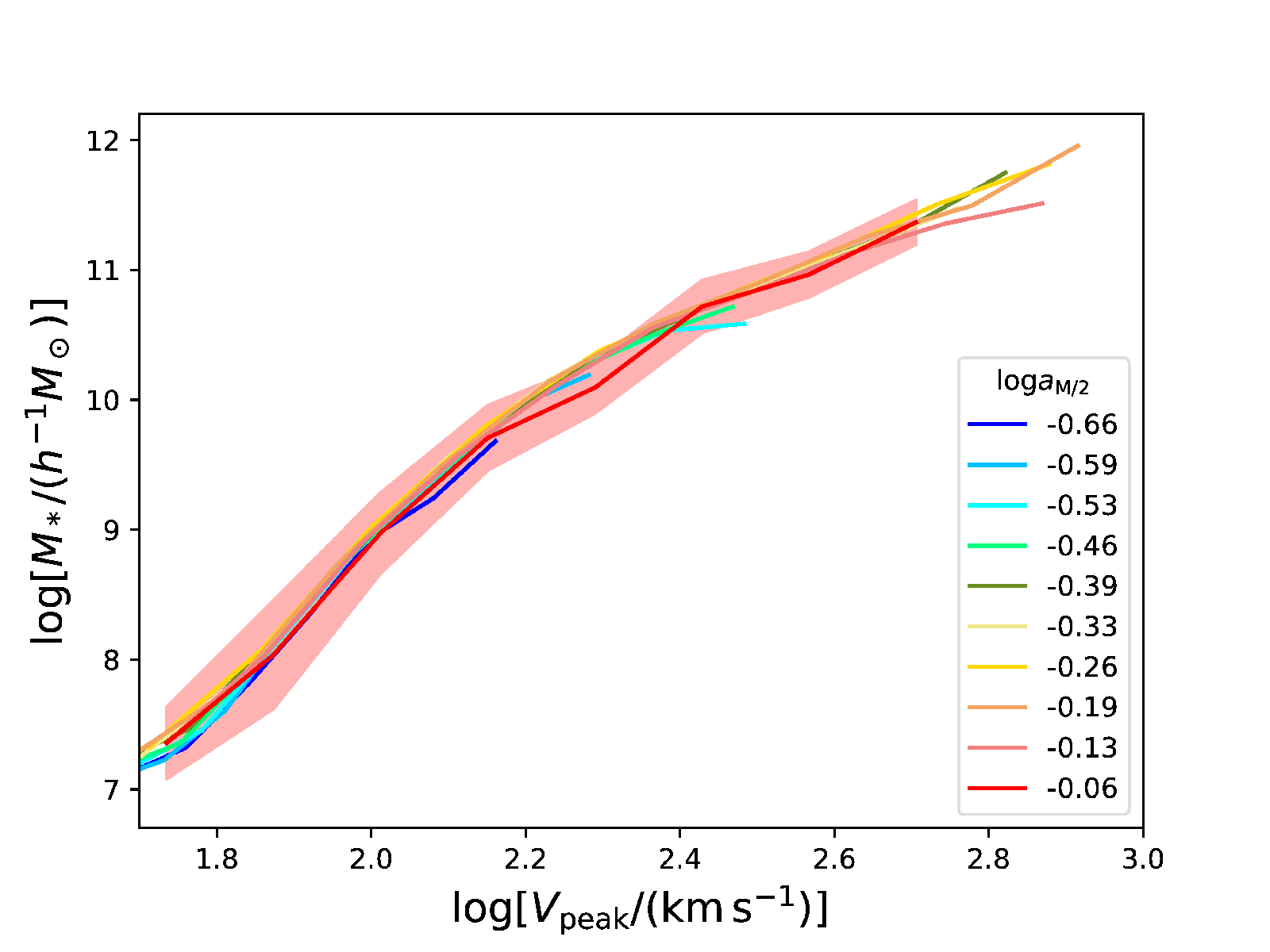}
	\end{subfigure}
	\hfill
\caption{
Top-left: $\Mstar$ as function of $\Mh$ for central galaxies. The galaxies are colour-coded according to $\log[\Vp/({\rm km\, s^{-1}})]$. For galaxies in each bin of $\log\Vp$, the contours correspond to the 68.3 and 95.4 per cent distribution, respectively. 
Top-right: $\Mstar$ as function of \Vpeak\ for central galaxies, colour coded according to  $\log[\Mh/(\hinvMsun)]$.
Bottom-left: $\Mstar$ as function of $\Mh$ for central galaxies, with the mean relation colour-coded according to the values of \ahalf. For clarity, the scatter in the mean relation is only shown for the bin with the highest \ahalf\ (latest forming haloes). 
Bottom-right: $\Mstar$ as function of \Vpeak\ for central galaxies, colour-coded according to \ahalf, with the shaded region illustrating the scatter for the bin with the highest \ahalf. Note the remarkable result that $\Mstar$ does not depends on \ahalf\ at fixed $\Vp$ (compared to the $\Mh$ case in the bottom-left panel).
}
\label{fig:mstar_plot}
\end{figure*}

\begin{figure}
    \includegraphics[width=0.48\textwidth]{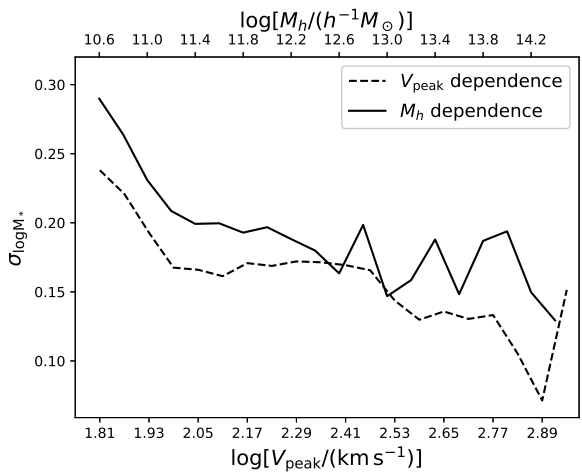}
\caption{
Standard deviation in $\log\Mstar$ as a function of $\Mh$ (solid) and \Vpeak\ (dashed). The correspondence between $\Mh$ and \Vpeak\ is from the mean relation \Vpeak$\propto \Mh^{1/3}$ in equation~(\ref{eq:VM}). Poisson errors in $\log\Mstar$ caused by finite number of star particles have been subtracted in quadrature.}
\label{fig:mstarscatter}
\end{figure}

\begin{figure*}
	\centering
	\begin{subfigure}[h]{0.48\textwidth}
		\includegraphics[width=\textwidth]{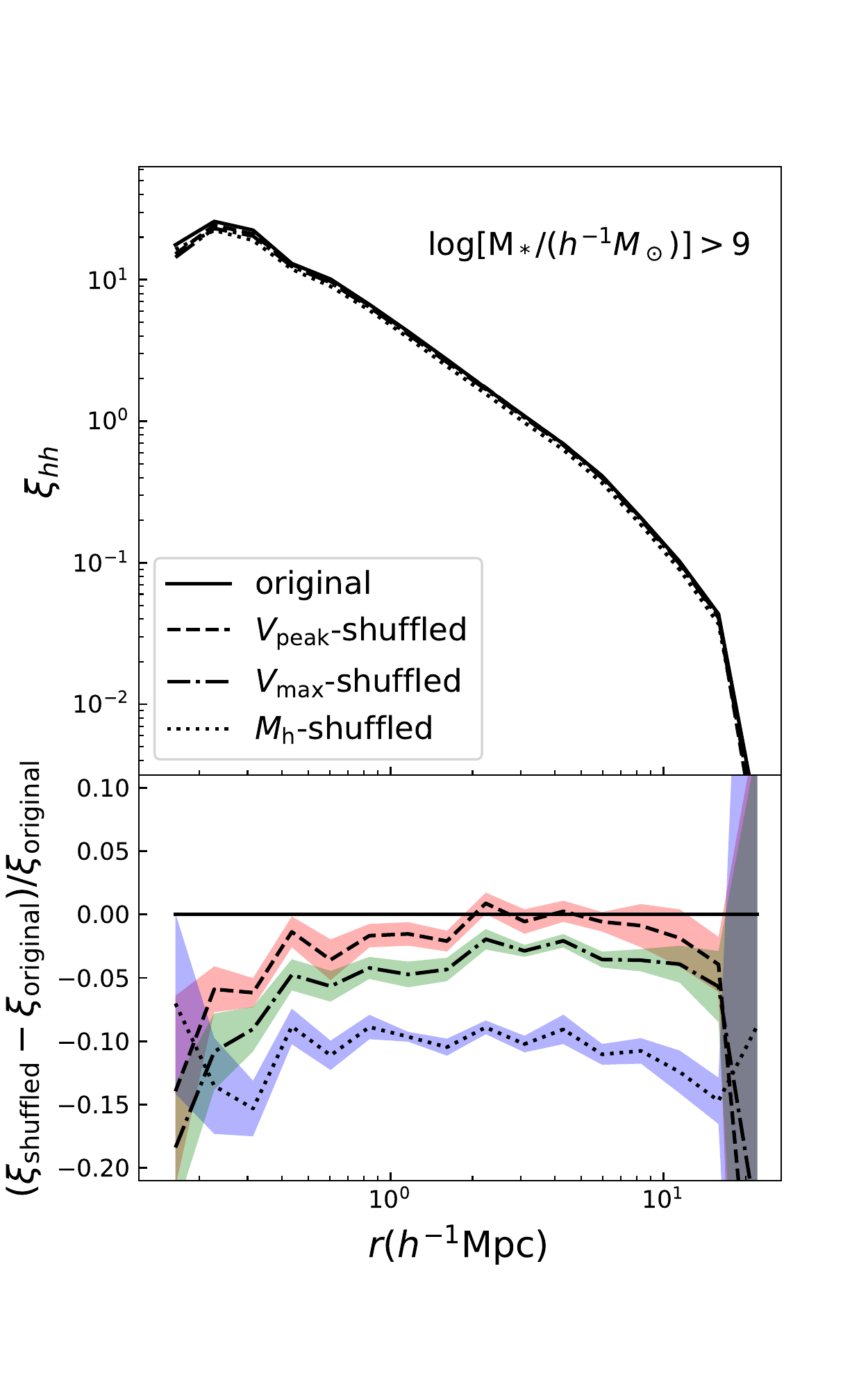}
	\end{subfigure}
	\hfill
	\begin{subfigure}[h]{0.48\textwidth}
        \includegraphics[width=\textwidth]{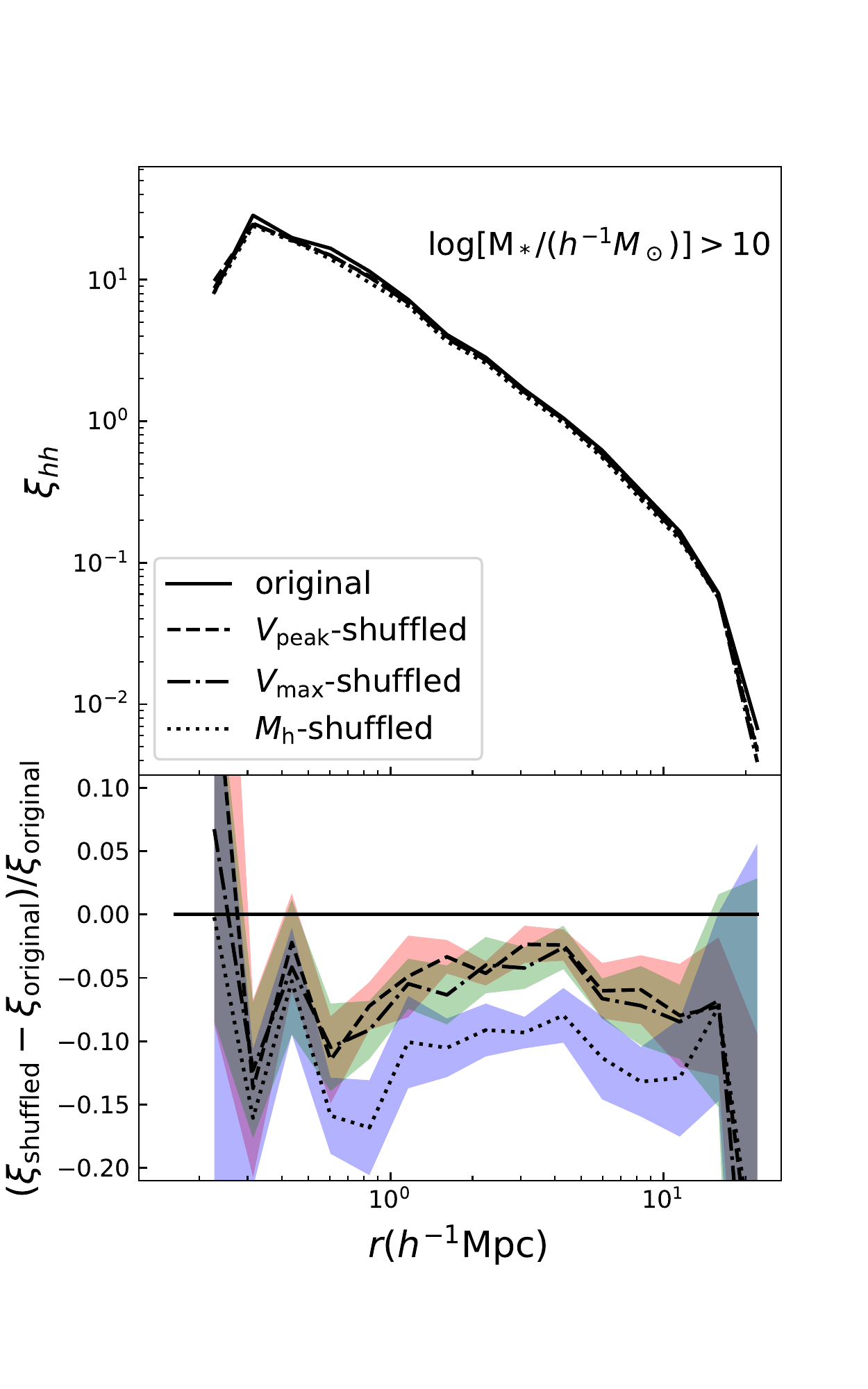}
	\end{subfigure}
	\hfill
\caption{Comparison of two-point correlation functions of original and shuffled central galaxy samples, selected by stellar mass thresholds. Left: The top panel compares the two-point correlation functions of central galaxies with $\log[\Mstar/(\hinvMsun)]$>9 from the original, unshuffled sample (solid) and the $\Mh$/\Vmax/\Vpeak-shuffled samples (dotted/dot-dashed/dashed). The bottom panel shows fractional differences between those from the shuffled and original samples, with each shaded band indicating the uncertainty estimated from 10 shuffling realisations. Right: same as the left, but for central galaxy samples with $\log[\Mstar/(\hinvMsun)]$>10. See details in section~\ref{sec:mstardep}.
}
\label{fig:xihh}
\end{figure*}

Galaxy stellar mass is a primary property inferred from observation. The relation between this primary galaxy property and certain primary halo property (e.g., $\Mh$ and \Vpeak) can be established based on subhalo abundance matching or modelling the stellar mass dependent clustering, which encodes information about galaxy formation. Here we show the relation predicted by the Illustris simulation and study how tight stellar mass correlates with $\Mh$ and \Vpeak.

The top-left panel of Fig.~\ref{fig:mstar_plot} shows $\Mstar$ as function of $\Mh$, colour-coded with values of $\Vp$.  Galaxy stellar mass $\Mstar$ increases steeply with $\Mh$ at $\log[\Mstar/(\hinvMsun)]<12$ and then slowly at $\log[\Mstar/(\hinvMsun)]>12$, a trend similar to that inferred from observation \citep[e.g.][]{Behroozi10,Leauthaud12,Zu15}. The scatter in $\Mstar$ at fixed halo mass in the $\Mstar$--$\Mh$ relation decreases with increasing $\Mh$ (solid curve in Fig.~\ref{fig:mstarscatter}), varying from about 0.3 dex at the low-mass end to about 0.17 dex at the high-mass end. The scatter at the high-mass end is consistent with the value $\sim 0.16$ dex inferred from galaxy clustering modelling \citep[e.g.][]{Tinker17a}. One source of the scatter can be the halo formation history \citep{Tinker17b}, which may affect the growth history of stellar mass (either from star formation or galaxy merging; e.g. \citealt{Gu16}). 

The colour code in $\Vp$ in the top-left panel enables us to see how the scatter in the $\Mstar$--$\Mh$ relation may be connected to halo assembly. On average, \Vpeak\ and $\Mh$ are correlated, and the mean relation is found to be well described by
\begin{equation}
\label{eq:VM}
\Vp = 170 \left(\frac{\Mh}{10^{12}\hinvMsun}\right)^{1/3}\, {\rm km\, s^{-1}}
\end{equation}
in the Illustris simulation. However, there is scatter on top of the mean relation, and at fixed $\Mh$, the distribution of \Vpeak\ reflects that in the assembly history. As can be seen in the top-left panel, the assembly of haloes encoded in \Vpeak\ does contribute to the scatter in the $\Mstar$--$\Mh$ relation -- at fixed $\Mh$, galaxies residing in haloes of higher \Vpeak\ tend to have higher stellar mass, especially at the low mass end.

To see how well the scatter in $\Mstar$ can be attributed to the scatter in \Vpeak, in the top-right panel of Fig.~\ref{fig:mstar_plot}, we plot the $\Mstar$--\Vpeak\ relation. It follows a similar trend seen in the $\Mstar$--$\Mh$ relation, steeper (shallower) dependence of $\Mstar$ on \Vpeak\ at the low (high) \Vpeak\ end, which is expected given the correlation between \Vpeak\ and $\Mh$. The $\Mstar$--\Vpeak\ relation appears to be tighter than the $\Mstar$--$\Mh$ relation, in the sense that at fixed \Vpeak\ the scatter in $\Mstar$ is lower than that at the corresponding $\Mh$ (see \citealt{Matthee17} for a similar result in terms of $z=0$ maximum halo circular velocity with the EAGLE simulation). The scatter varies from $\sim$0.28 dex at low \Vpeak\ to $\sim$0.13 dex at high \Vpeak\ (dashed curve in  Fig.~\ref{fig:mstarscatter}). In the $\Mstar$--\Vpeak\ plot (top-right panel), the contours are colour-coded by $\Mh$. Unlike the $\Mh$ case in the top-left panel, we find that $\Mstar$ does not show a clear dependence on $\Mh$ -- at fixed $\Vp$ (i.e. by taking a vertical cut in the plot), $\Mstar$ in haloes of different $\Mh$ appears to follow similar mean and scatter. 

In the bottom-left panel, we show the effect of the other halo assembly property \ahalf\ on the $\Mstar$--$\Mh$ relation. Each curve show the mean $\Mstar$--$\Mh$ relation for haloes in one \ahalf\ bin. The scatter around the mean relation is illustrated with the shaded region, only shown for the latest forming haloes to avoid crowdedness. There is a clear and substantial dependence of the $\Mstar$--$\Mh$ relation on the assembly property \ahalf -- at fixed $\Mh$, haloes forming earlier (smaller \ahalf) tend to host galaxies of higher $\Mstar$. Such a trend is consistent with previous work based on the EAGLE simulation \citep{Matthee17} and semi-analytic galaxy formation model \citep{Zehavi18}.

Switching to the $\Mstar$--$\Vp$ relation (bottom-right panel), we find that at fixed $\Vp$ there is no dependence of $\Mstar$ on the assembly property \ahalf\ , with the curves of different \ahalf\ bins all falling on top of each other. As further shown in section~\ref{sec:ghrelation}, $\Vp$ is able to absorb the effect on $\Mstar$ from any other halo assembly variable. This is consistent with the results using $z=0$ maximum halo circular velocity \citep{Matthee17}. We will discuss this remarkable result in section~\ref{sec:summary}. 

The other commonly adopted halo circular velocity quantity is \Vmax, the present-day maximum circular velocity of a halo. In Appendix~\ref{sec:appendixa}, we show that the $\Mstar$--\Vmax\ relation is similar to the $\Mstar$--\Vpeak\ relation, but with the former showing a little bit residual dependence on halo formation time. In Appendix~\ref{sec:appendixa}, by replacing \Vpeak\ used here from the full physics simulation with the one from the dark-matter-only simulation, we also show that baryon effect on the halo potential plays almost no role in driving \Vpeak\ to track $\Mstar$.

The tighter correlation between $\Mstar$--\Vpeak, in comparison to $\Mstar$--$\Mh$, suggests that in galaxy clustering modelling galaxy assembly bias effect can be partially account for by switching from a $\Mh$--based model to a \Vpeak-based model. That is, the galaxy-halo relation is parameterised as a function of \Vpeak. This is in line with the finding by \citet{Chaves-Montero16} using the EAGLE simulation, while their velocity quantity most strongly correlating with $\Mstar$ is slightly different, ${\rm V_{relax}}$, the highest value of the maximum circular velocity of a subhalo with a relaxation criterion imposed. \citet{Chaves-Montero16} show that ${\rm V_{relax}}$ is able to capture the majority (but not all) of the galaxy assembly bias effect on galaxy clustering for stellar-mass-based samples. The tighter correlation between $\Mstar$--\Vpeak\ may also be the reason that when using subhalo abundance matching or its variants to model galaxy clustering, halo circular velocity-based models usually have good performances \citep[e.g.][]{Reddick13,Guo16}.

To investigate how well assembly bias effect in Illustris central galaxy clustering can be absorbed by \Vpeak, we perform a test by constructing shuffled galaxy catalogues \citep[e.g.][]{Croton07,Zu08} and compare the two-point correlation functions of the original and shuffled central galaxy samples, selected by stellar mass thresholds. By dividing haloes into small \Vpeak\ bins, we randomly shuffle the values of $\Mstar$ among haloes in each \Vpeak\ bin, which eliminates the dependence of $\Mstar$ on halo properties other than \Vpeak. If \Vpeak\ can account for all assembly effects on central galaxy $\Mstar$, we would expect the \Vpeak-shuffled sample to have the same two-point correlation function as the original, unshuffled sample. For comparison, we also similarly construct \Vmax-shuffled and $\Mh$-shuffled samples. In the left panel of Fig.~\ref{fig:xihh}, the result is shown for the $\log[\Mstar/(\hinvMsun)]$>9 sample. The $\Mh$-shuffled sample show the largest deviation in the two-point correlation function from the original sample, with the assembly effect on galaxy clustering at a level of 10 per cent. The level of deviation is small (below $\sim$2 per cent on most scales above 1$\hinvMpc$) for the \Vpeak-shuffled sample, suggesting that \Vpeak\ is able to absorb the majority of the assembly bias on stellar mass. The result with the \Vmax-shuffled sample is in between, and \Vmax\ is not as efficient as \Vpeak\ in absorbing the assembly bias effect on stellar mass (consistent with that seen in Appendix~\ref{sec:appendixa}). For the $\log[\Mstar/(\hinvMsun)]$>10 sample (right panel), the assembly bias effect is also at the 10 per cent level, and both \Vpeak\ and \Vmax\ are able to roughly absorb more than half of it, with \Vpeak\ behaving marginally better than \Vmax. We note that the overall trend is similar to that in \citet{Zehavi19} based on galaxies in a semi-analytic galaxy formation model (top panel of their fig.8 for central galaxies). As a whole, we see that \Vpeak\ is able to absorb a large fraction of the assembly effect on central galaxy stellar mass, more efficient for lower mass haloes, suggesting that \Vpeak\ may serve as a good halo variable to replace $\Mh$ in modelling clustering of galaxies selected by stellar mass. 

In what follows, besides a further investigation of the dependence of stellar mass on halo assembly variables, we also extend the study to other galaxy properties.

\subsection{Relationship between galaxy properties and halo properties}
\label{sec:ghrelation}

\begin{figure*}
    \includegraphics[width=1.1\textwidth]{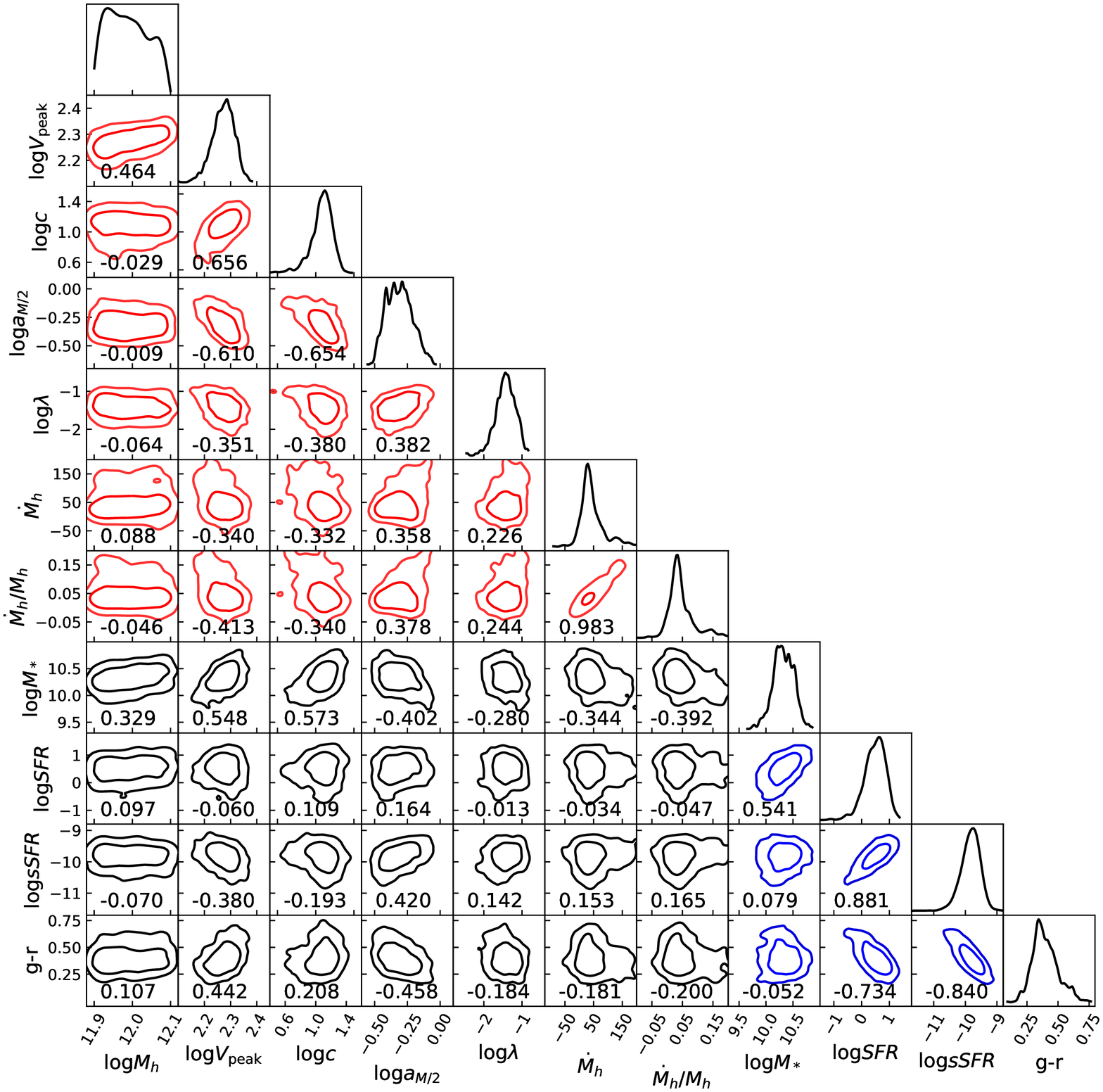}
\caption{
Relation between each pair of galaxy and/or halo properties at $\log[\Mh/(\hinvMsun)]\sim 12$. In each contour panel, the two contours show the central 68.3 and 95.4 per cent of the distribution of the pair of properties. The panels with red contours (i.e. the top 6 rows and left 6 columns of contour panels) display the correlations between halo properties. Those with black contours (i.e. the bottom 4 rows and the left 7 columns of contour panels) show the correlations between galaxy and halo properties, and those with blue contours (i.e. the right 3 columns of contour panels) are for the correlations between pairs of galaxy properties. The number in each contour panel is the Pearson correlation coefficient for the pair of properties. The histogram at the top panel of each column is the probability distribution function of the variable of that column.
}
\label{fig:gh_m}
\end{figure*}

\begin{figure*}
	 \includegraphics[width=1.1\textwidth]{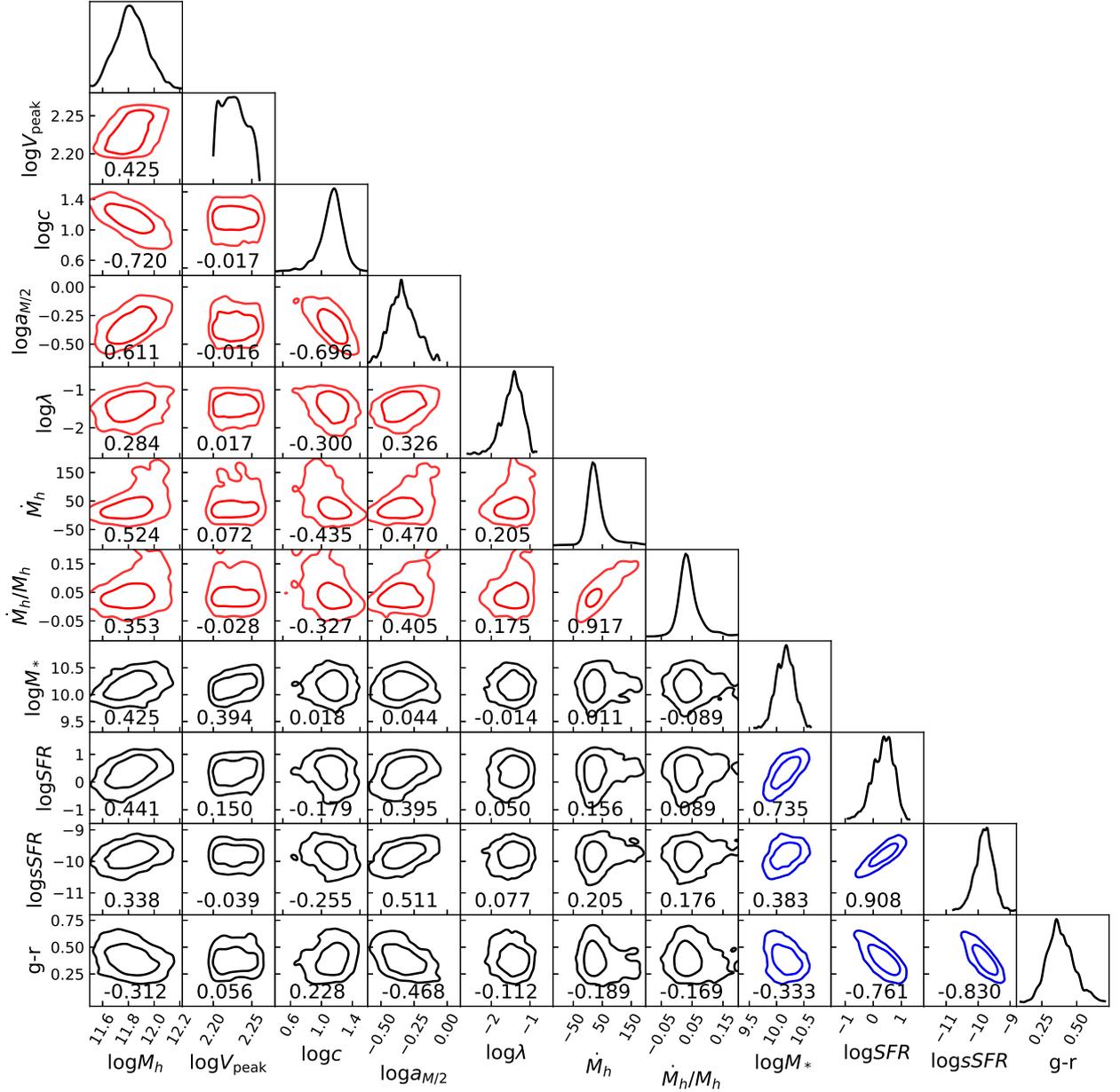}
\caption{
Same as Fig.~\ref{fig:gh_m}, but at fixed $\log[{\rm V_{peak}/(km\, s^{-1})}]\sim 2.23$. Note particularly the lack of correlation of $\Mstar$ with other assembly variables (including $c$, \ahalf, $\lambda$, $\Mdot$, $\Mdot/\Mh$), in contrast with the case in Fig.~\ref{fig:gh_m}.
}
\label{fig:gh_vp}
\end{figure*}

\begin{figure*}
	\centering
	\includegraphics[width=1.1\textwidth]{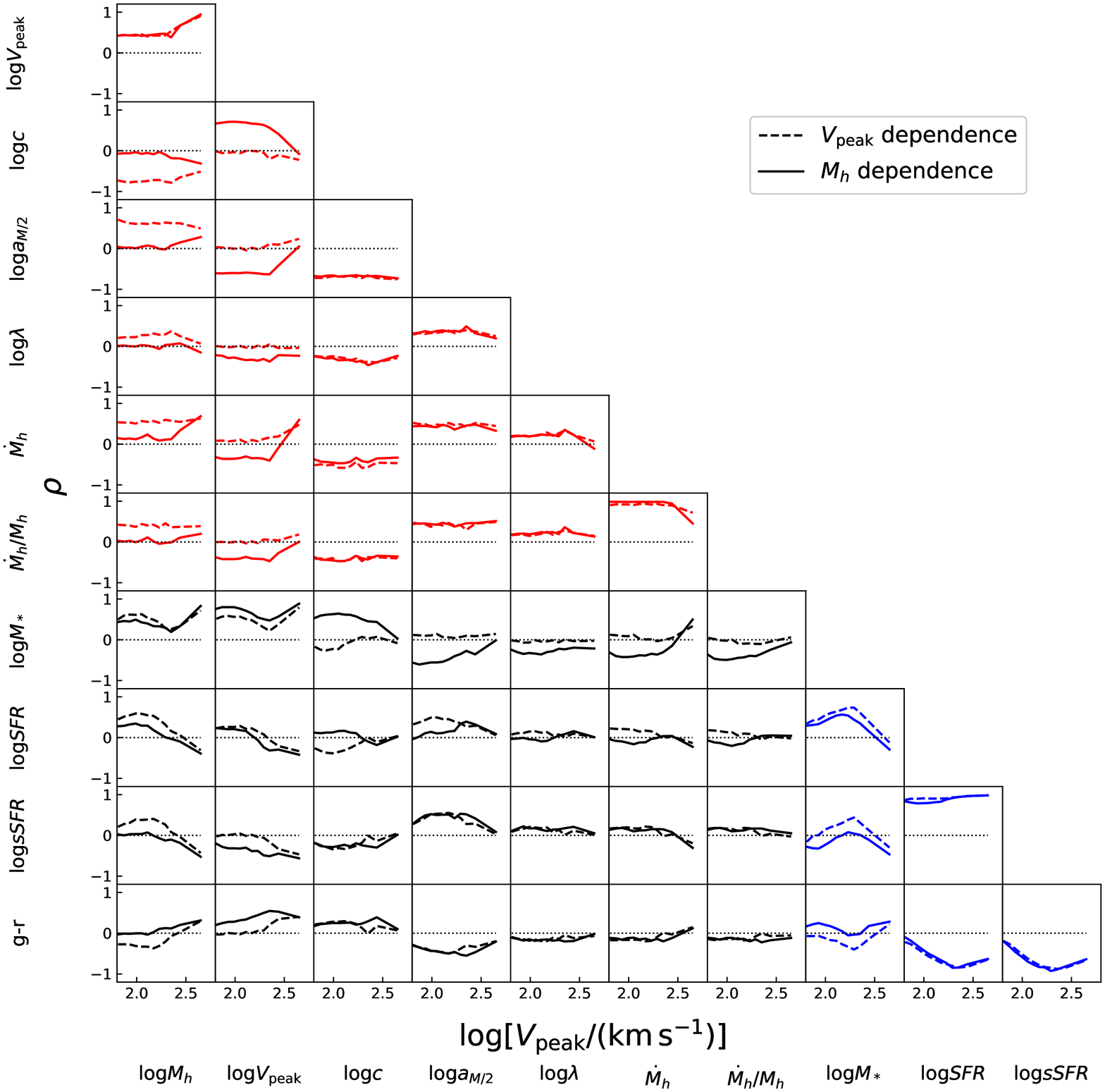}
\caption{
Pearson correlation coefficient $\rho$ of each pair of galaxy and/or halo properties as a function of $\Mh$ (solid) and $\Vp$ (dashed). 
The galaxy and halo properties are marked to the far left of each row and at the bottom of each column. For clarity, we only label the values of $\Vp$ on the horizontal axis, and the values of $\Mh$ can be inferred from $\Mh\propto \Vp^3$ from equation~(\ref{eq:VM}). As with Fig.~\ref{fig:gh_m}, panels with red, black, and blue curves are for correlations between halo-halo, galaxy-halo, and galaxy-galaxy properties, respectively. In each panel, the dotted horizontal line indicates no correlation. Note particularly the lack of correlation of $\Mstar$ with other assembly variables (including $c$, \ahalf, $\lambda$, $\Mdot$, $\Mdot/\Mh$) for the $\Vp$ dependence case, in contrast with the $\Mh$ dependent case. Also the correlations of SFR, sSFR, and colour with halo assembly variables  show more consistent behaviours in the $\Vp$ dependence case.
}
\label{fig:gh_pearson}
\end{figure*}

With the 7 halo properties ($\Mh$, \Vpeak, $c$, \ahalf, $\lambda$, $\Mdot$, and $\Mdot/\Mh$) and the 4 central galaxy properties ($\Mstar$, SFR, sSFR, and $g-r$ colour), we study the correlations between them. To aid the discussion, we also present the correlations among halo properties and those among galaxy properties. 

\subsubsection{At fixed $\Mh$}
Fig.~\ref{fig:gh_m} shows the correlation between each pair of the halo and galaxy properties for central galaxies in haloes of a narrow mass bin, $\log[\Mh/(\hinvMsun)]=12.0\pm 0.1$. In each contour panel, the contours indicate the 68.3 and 95.4 per cent of the distribution of the pair of properties. The number labelled in each panel is the Pearson correlation coefficient $\rho$ of the two properties, indicating how strong the correlation is. It is calculated as 
\begin{equation}
\rho = \frac{\langle xy \rangle-\langle x \rangle \langle y \rangle}{\sigma_x\sigma_y},
\end{equation}
where $x$ and $y$ denote the two properties, $\langle \rangle$ means average, and $\sigma_x$ and $\sigma_y$ are the standard deviations of $x$ and $y$. The panel at the top of each column shows the marginalised distribution of the property labelled at the $x$-axis of the column.

The panels with red contours (i.e. the top 6 rows and left 6 columns of contour panels) display the correlation between halo properties. Within the small but finite halo mass bin, all but one halo property shows almost no correlation with $\Mh$ (correlation coefficient close to zero). The exception is $\Vp$, and the correlation is simply driven by the $\Vp\propto\Mh^{1/3}$ mean relation. At fixed $\Mh$, any pair of halo properties show a substantial correlation (with $|\rho|$ above 0.2). The nearly perfect correlation ($\rho=0.983$) between $\Mdot$ and $\Mdot/\Mh$ is a consequence of fixed $\Mh$. Overall the correlation trend is that haloes of higher $\Vp$ are more concentrated, form earlier, spin more slowly, and have lower accretion rate, which have been seen in previous work \citep[e.g.][]{Jeeson11,Han19,Xu18}.

The panels with black contours (i.e. the bottom 4 rows and the left 7 columns of contour panels) show the correlation between halo and galaxy properties. The correlation between $\Mstar$ and $\Mh$ shows up because of the finite size of the halo mass bin. At fixed $\Mh$, the central galaxy stellar mass $\Mstar$ correlates with all other halo properties -- haloes of higher $\Vp$, higher concentration, earlier formation, lower spin, and lower accretion rate tend to host more massive central galaxies. On the contrary, the SFR shows no strong correlation with any halo properties. The most significant one is with halo formation time ($\rho\sim 0.16$), with on average higher SFR in haloes of later formation. Given the substantial  correlation between $\Mstar$ and halo properties and the weak or lack of correlation between SFR and halo properties, the sSFR ($\equiv {\rm SFR}/\Mstar$) is expected to correlate well with halo properties, but in an trend opposite to and weaker than that with $\Mstar$. This is indeed the case. The most significant correlation is with \Vpeak\ or \ahalf\ (both with $|\rho|\sim 0.4$). The correlation between sSFR and the average halo accretion rate over the past dynamic time is there but not strong ($\rho\sim 0.16$). The correlation between galaxy colour $g-r$ and halo properties essentially follows the case of sSFR (with a sign change in $\rho$; redder galaxies having lower sSFR).

The panels with blue contours (i.e. the right 3 columns of contour panels), the correlations between pairs of galaxy properties at fixed halo mass are shown. The SFR positively correlates with $\Mstar$ ($\rho\sim 0.54$), and the mean relation has a slope close to unity, ${\rm SFR} \propto \Mstar$. It resembles the star-forming main sequence \citep[e.g.][]{Brinchman04,Speagle14,Santini17}. That is, even if we only consider central galaxies in haloes of fixed mass, the star-forming main sequence emerges. Note that this SFR--$\Mstar$ correlation is not driven by the correlation of SFR and $\Mstar$ with a common halo variable we consider here. In fact, from Fig.~\ref{fig:gh_m}, it can be seen that their correlation with a common halo variable may lead to the opposite effect. For example, haloes of earlier formation tend to host central galaxies of higher $\Mstar$ and lower SFR, and naively this would imply an anti-correlation between SFR and $\Mstar$, opposite to what is found here. While it is possible that the halo-level star-forming main sequence is related to a halo variable not considered here, it is more likely that the sequence is driven by baryonic physics, which may have complicated dependence on or decouple from halo formation history. Unlike the SFR, the sSFR shows little dependence on $\Mstar$. However, the sSFR is tightly correlated with the SFR ($\rho\sim 0.88$). Given that sSFR$\equiv$SFR/$\Mstar$, the pattern in the mutual correlations among $\Mstar$, SFR, and sSFR can be achieved if the SFR--$\Mstar$ correlation coefficient is close to the ratio of the scatters in $\log\Mstar$ and $\log{\rm SFR}$\footnote{To see this, let $x=\log\Mstar$, $y=\log{\rm SFR}$, and $z=\log{\rm sSFR}=y-x$. We can derive the relation among the correlation coefficients, $\rho_{xz}/\rho_{yz}=(\rho_{xy}-\sigma_x/\sigma_y)/(1-\rho_{xy}\sigma_x/\sigma_y)$. For $\rho_{xz}$ to be near zero, we have $\rho_{xy}\sim \sigma_x/\sigma_y$. }, which appears to be the case. Galaxy $g-r$ colour strongly correlates with sSFR and follows the same trends as sSFR in its correlations with $\Mstar$ and SFR.

Overall, at fixed halo mass, galaxy properties other than SFR show significant correlations with one or more halo properties, manifesting galaxy assembly bias at the level of haloes.  The correlations among galaxy properties, however, may largely result from baryonic physics, given that the trend cannot be simply explained by their correlation with halo properties. In section~\ref{sec:mstardep}, it is found that switching from $\Mh$ to $\Vp$ can remove the dependence of $\Mstar$ on other halo properties. We now extend the investigation to other galaxy properties.

\subsubsection{At fixed \Vpeak}

Fig.~\ref{fig:gh_vp} is similar to Fig.~\ref{fig:gh_m}, but the correlations are presented for haloes at a fixed $\Vp$ bin, $\log[\Vp/({\rm km\, s^{-1}})]=2.23\pm 0.03$. The correlations among halo properties (in panels with red contours) are similar to those in Fig.~\ref{fig:gh_m}, and there are additional correlations between $\Mh$ and other halo properties. 

The galaxy-halo correlations are shown in panels with black contours. The finite bin size in $\Vp$ makes the $\Mstar$--$\Vp$ correlation show up. Other than this (and the one with $\Mh$), $\Mstar$ does not correlate with any other halo properties at fixed $\Vp$, reinforcing the result in section~\ref{sec:mstardep}. It indicates that the correlations of $\Mstar$ with halo properties seen at fixed $\Mh$ (Fig.~\ref{fig:gh_m}) can be attributed to the $\Mstar$--$\Vp$ correlation and the correlation of $\Vp$ with other halo properties. For the SFR, at fixed $\Vp$, it correlates significantly with $\Mh$ and \ahalf, higher SFR in haloes of higher mass and later formation. The correlations between SFR and other halo properties are weak. The correlations between sSFR (or colour) and halo properties closely follow the SFR case. 

In terms of the galaxy-galaxy correlations, the trends are similar to those seen at fixed halo mass, except that the sSFR and colour now show clear correlations with $\Mstar$. 

As a whole, using $\Vp$ as the halo variable largely removes the correlations between $\Mstar$ and other halo assembly properties, and the dependences of SFR, sSFR, and colour on halo assembly variables follow each other. 

\subsubsection{Dependence on $\Mh$ and \Vpeak}

The correlations shown in Fig.~\ref{fig:gh_m} and Fig.~\ref{fig:gh_vp} are for haloes of $\log[\Mh/(\hinvMsun)]\sim 12.0$ and $\log[\Vp/({\rm km\, s^{-1}})] \sim 2.23$. To obtain a full picture, in Fig.~\ref{fig:gh_pearson} we present the $\Mh$ and $\Vp$ dependent Pearson correlation coefficients for the various pairs of galaxy and halo properties, by performing the calculation in different $\Mh$ and $\Vp$ bins, respectively. The panels correspond to those in Fig.~\ref{fig:gh_m} and Fig.~\ref{fig:gh_vp}, and the correlation shown in a panel of a given row and column is between the property as labelled at the far left of the row and that at the bottom of the column. In each panel, the solid (dashed) curve is the dependence of $\rho$ on $\Mh$ ($\Vp$), with zero correlation marked by the black dotted curve. Note that only $\Vp$ is shown on the $x$-axis and the corresponding $\Mh$ can be obtained according to $\Mh\propto \Vp^3$ from equation~(\ref{eq:VM}).

The panels with red curves show the correlations among halo properties. If we limit to halo properties other than $\Mh$ and $\Vp$, we find that the correlation of any pair of the assembly variables only weakly depends on $\Mh$ or $\Vp$ if any (manifested by the nearly flat curves) and that the correlation strength does not depend on whether we use $\Mh$ or $\Vp$ bins (manifested by the highly overlapped solid and dashed curves). 

For the galaxy-halo correlations (in panels with black curves), 
in terms of $\Mh$ dependence, the strongest correlation between galaxy and halo property is found between $\Mstar$ and $\Vp$/$c$/\ahalf\ in low mass halos, with $|\rho|\sim 0.5$--0.6. 
It holds true in the full range of haloes considered here that using $\Vp$ largely removes the correlation between $\Mstar$ and any other halo assembly variable (dashed curves around zero). The only exception is that $\Mstar$ appears to be slightly anti-correlated with $c$ in low-$\Vp$ haloes. With $\Vp$ the dependences of correlation on $\Vp$ for SFR, sSFR, and colour closely track each other, which is not the case for those on $\Mh$. With $\Vp$ as the primary halo variable, star formation related properties (SFR, sSFR, and colour) mainly show dependence on halo formation time and then halo concentration.

For galaxy properties (in panels with blue curves), the correlation between SFR and $\Mstar$ reaches a maximum in haloes of $\sim 10^{12}\hinvMsun$. It weakens in haloes of higher $\Mh$ or $\Vp$, probably because galaxies move away from the star-forming main sequence and passive evolution starts to dominate. However, the tight sSFR--SFR correlation persists over the full $\Mh$ or $\Vp$ range. For colour, the correlation with SFR and sSFR is weak in low-$\Mh$ or low-$\Vp$ haloes and becomes stronger in haloes of higher $\Mh$ or $\Vp$. 

The results indicate that in the Illustris simulation galaxy formation ties to $\Vp$ more closely than $\Mh$. In comparison with the $\Mh$-based results, we find that in terms of $\Vp$, galaxy properties show a cleaner trend in the correlation with other halo assembly variable, such as the lack of correlation for $\Mstar$ and the similar correlation pattern for SFR/sSFR/colour. It suggests that $\Vp$--based halo model would be a good choice for capturing galaxy assembly bias effect (e.g. with $\Mstar$-based galaxy samples) and for studying galaxy assembly bias (e.g. with SFR/sSFR/colour-based samples). With $\Vp$ as the primary halo variable in the model, halo formation time and concentration would be the main options for the secondary variable to describe the relation between halos and star formation related quantities, with the former preferable. 

\subsection{Assembly bias of central galaxies}
\label{sec:gab}

\begin{figure}
	\centering
	\begin{subfigure}[h]{0.48\textwidth}
	    \includegraphics[width=\textwidth]{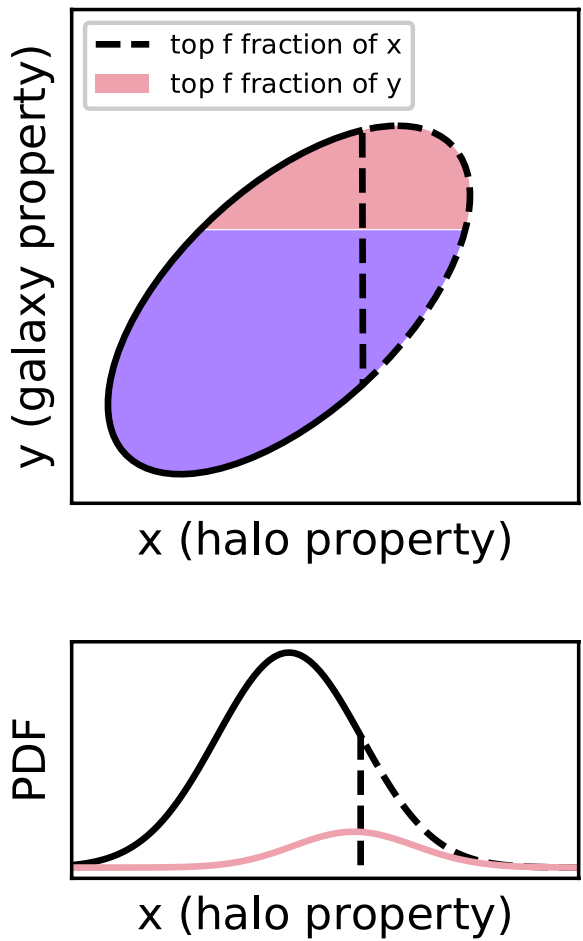}
    \end{subfigure}	    
	\hfill
\caption{
Illustration of the correlation between galaxy property and halo assembly property and the construction of the halo and galaxy samples for the study of galaxy assembly bias effect in section~\ref{sec:gab}. 
In the top panel, the ellipse denotes the joint distribution of halo property $x$ and galaxy property $y$ at fixed $\Mh$ or $\Vp$, which is assumed to follow a 2D Gaussian distribution. A halo sample is constructed with haloes of the top $f$ fraction of $x$ (indicated by the red region), and a galaxy sample is constructed with central galaxies of the top $f$ fraction of $y$ (indicated by the region inside the dashed curve). Shown in the bottom panel are the probability distribution functions of halo property $x$ for all the haloes at fixed $\Mh$ or $\Vp$ (black solid+dashed), the selected haloes (dashed), and the selected galaxies (red solid), respectively.
}
\label{fig:demo}
\end{figure}

\begin{figure*}
	\centering
	\begin{subfigure}[h]{0.48\textwidth}
		\includegraphics[width=\textwidth]{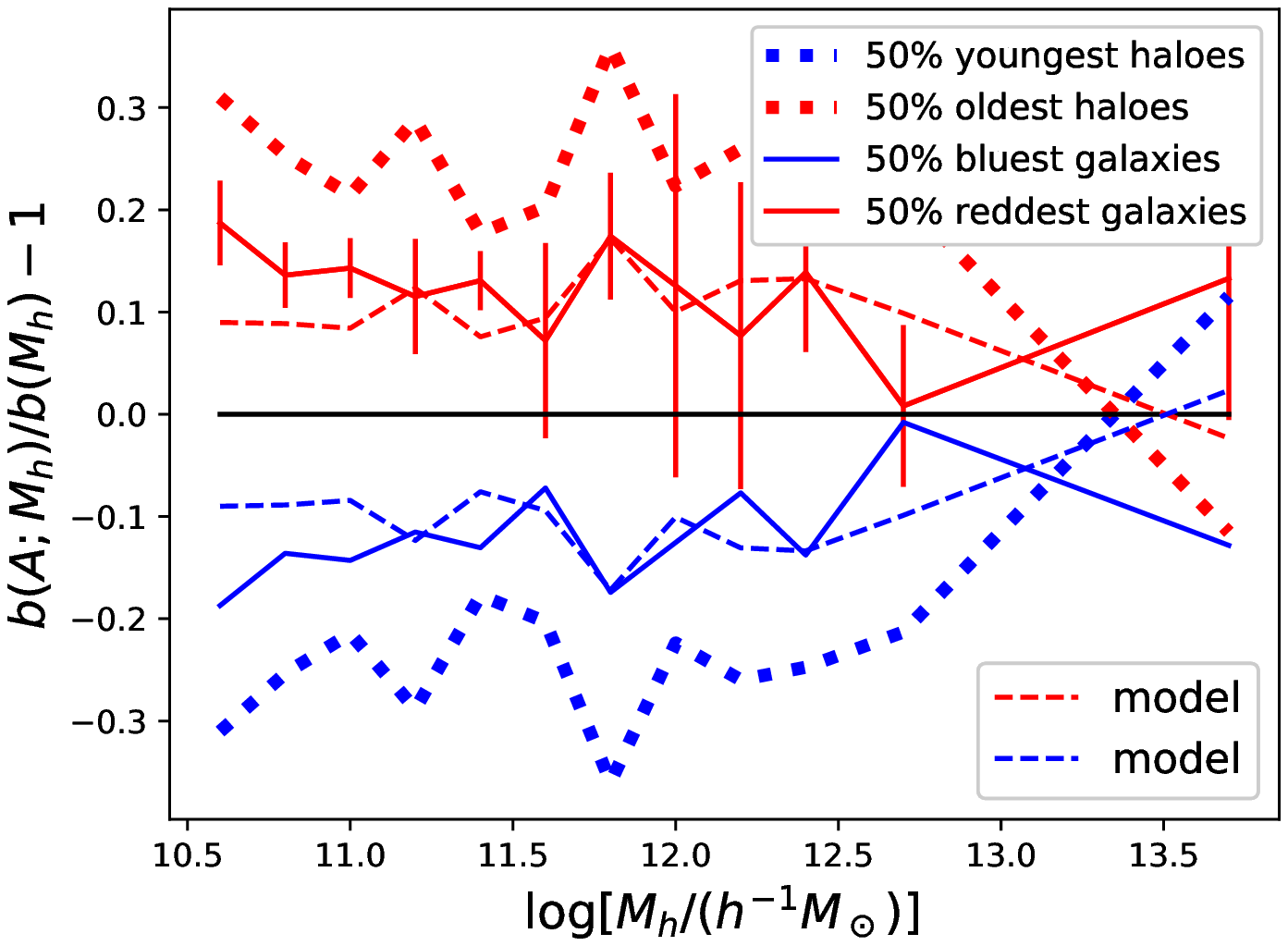}
	\end{subfigure}
	\hfill
	\begin{subfigure}[h]{0.48\textwidth}
                \includegraphics[width=\textwidth]{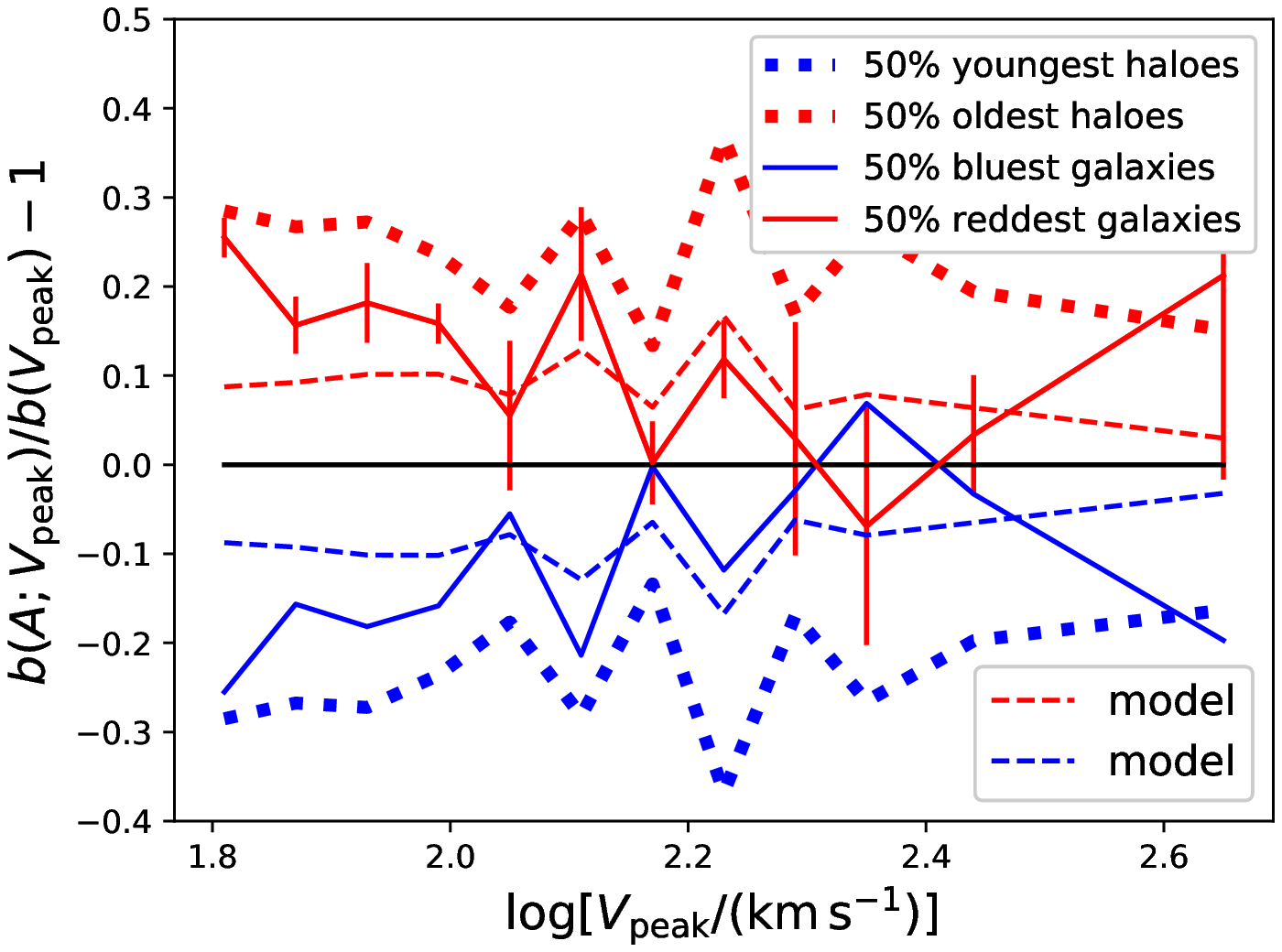}
	\end{subfigure}
	\hfill
	\begin{subfigure}[h]{0.48\textwidth}
                \includegraphics[width=\textwidth]{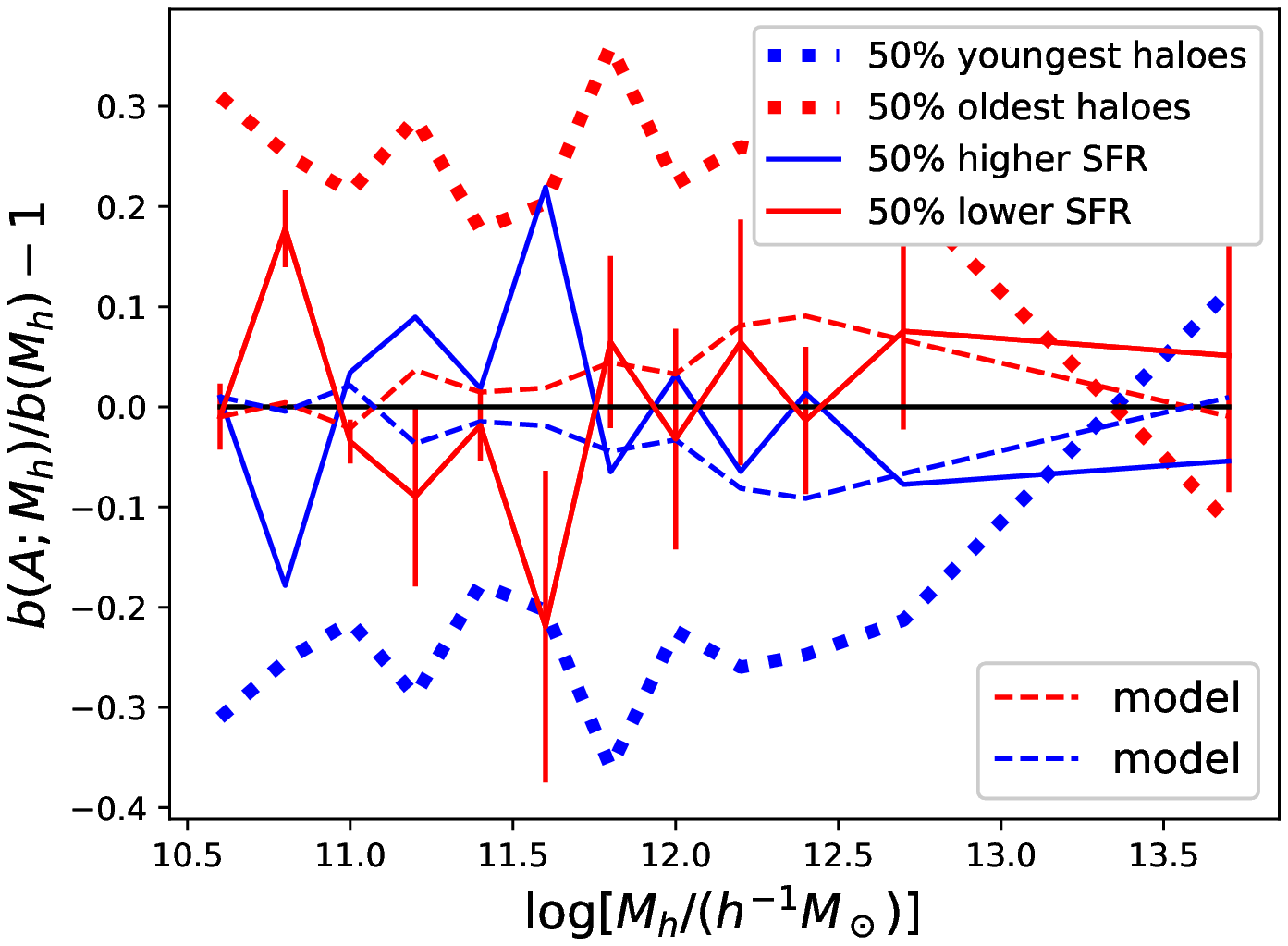}
	\end{subfigure}
	\hfill
	\begin{subfigure}[h]{0.48\textwidth}
		\includegraphics[width=\textwidth]{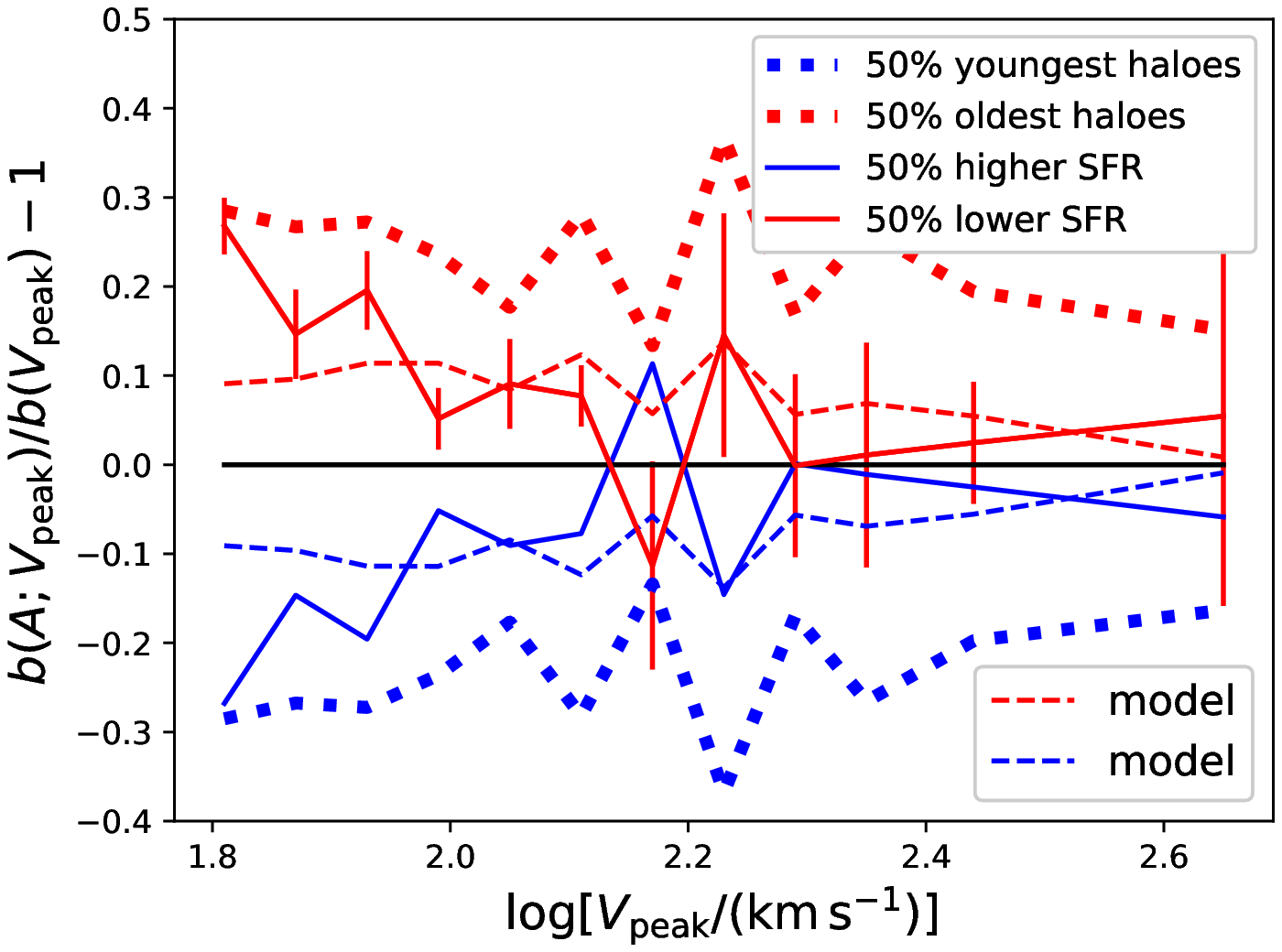}
	\end{subfigure}
	\hfill
	\begin{subfigure}[h]{0.48\textwidth}
                \includegraphics[width=\textwidth]{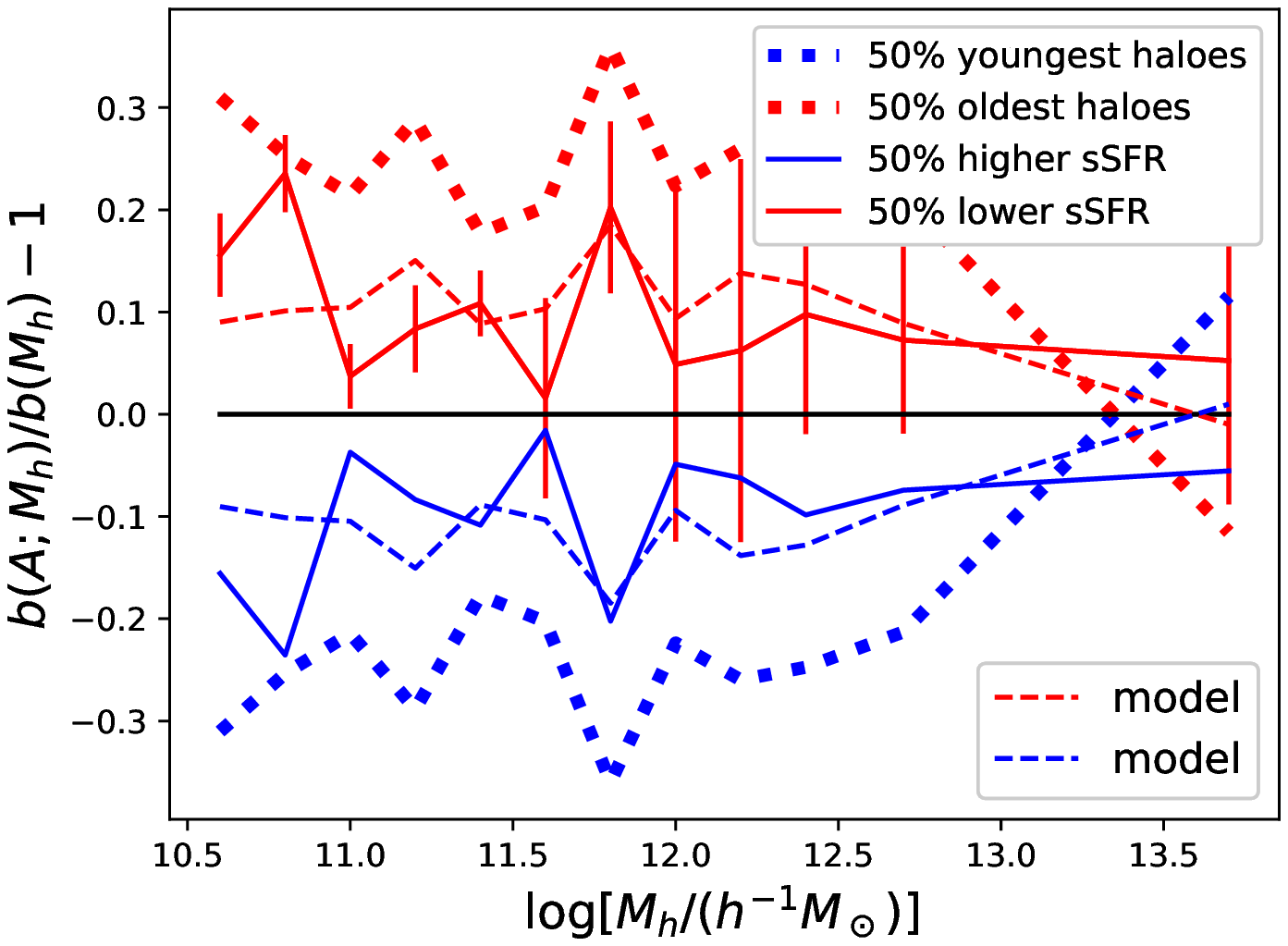}
	\end{subfigure}
	\hfill
	\begin{subfigure}[h]{0.48\textwidth}
                \includegraphics[width=\textwidth]{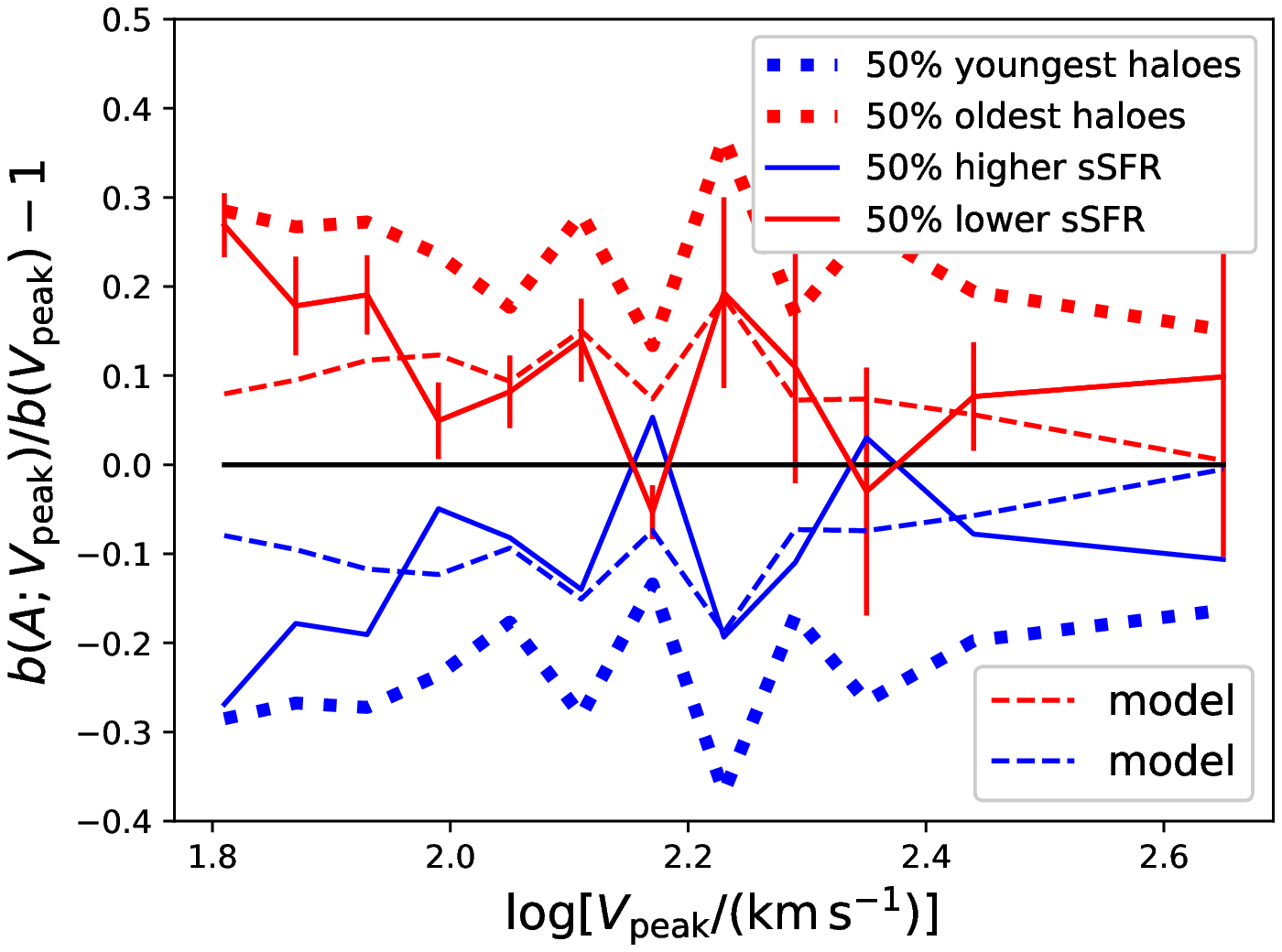}
	\end{subfigure}
	\hfill
\caption{
Connection between halo and galaxy assembly effect.
In the left panels, the quantities shown are the values of $b(A,\Mh)/b(\Mh)-1$ of different samples. For each sample, the quantity is the fractional difference of the bias factor of the sample selected based on property $A$ with respect to that of all haloes at fixed $\Mh$ (i.e. $\delta_h^b$ or $\delta_g^b$ defined in section~\ref{sec:gab}), which is used to characterise the magnitude of the assembly bias effect. The thick dotted curves are for halo samples selected based on halo formation time, and the solid curves are for central galaxy samples selected based on colour (top), SFR (middle), and sSFR (bottom). The thin dashed lines are the predictions from the simple model presented in section~\ref{sec:gab} according to the correlation between galaxy and halo properties ($\delta_g^b=\rho \delta_h^b$, with $\rho$ the correlation coefficient). 
For clarity, jackknife error bars are only shown for the solid curves in each panel. The right panels are the same, but for the assembly bias effect as a function of $\Vp$. See details in section~\ref{sec:gab}.
}
\label{fig:gabhab}
\end{figure*}

With the set of galaxy and halo properties investigated in section~\ref{sec:ghrelation}, we do not find a galaxy property that 100 per cent correlates with a halo assembly property. For the $\Mh$ dependence, the strongest correlation has $|\rho|\sim$0.5--0.6, between $\Mstar$ and $\Vp$/$c$/\ahalf. It means that halo assembly bias cannot be fully inherited by galaxies and be fully translated to galaxy assembly bias. Galaxy assembly bias should be different from halo assembly bias. For example, for haloes of the same mass, we can split them into two halo samples of low and high concentrations and then split central galaxies into two galaxy samples with low and high $\Mstar$. There would be a difference in the clustering of the two halo samples, as well as in that of the two galaxy samples. Given that $\Mstar$ is not perfectly correlated with $c$, we expect that the difference in the galaxy samples is smaller than that in the halo samples. As connecting galaxy assembly bias to halo assembly bias at the halo level can be an important ingredient in incorporating assembly bias effect into clustering model, we develop a simplified model below to understand the connection between galaxy and halo assembly bias.

Let us consider haloes at fixed $\Mh$ (or $\Vp$) and focus on one halo assembly variable $x$ (e.g. \ahalf\ or concentration) and one galaxy property $y$ (e.g. $\Mstar$ or SFR). Without losing generality, $y$ is assumed to be positively correlated with $x$. The joint distribution of $x$ and $y$ is illustrated by an ellipse in the top panel of Fig.~\ref{fig:demo}. We can form a halo sample by selecting the fraction $f$ of haloes with the highest $x$ (to the right of the vertical dashed line, the region inside the dashed curve) and a galaxy sample by selecting the same fraction of central galaxies with the highest $y$ (above the red-purple dividing line, the region in red). Then, what is the relation between the bias factors of the galaxy and halo sample? To answer the question, in addition to $p(x,y)$, we also need the dependence of the halo bias factor $b(x,y)$ on $x$ and $y$. For a sample $G$ defined by cuts in galaxy property $y$ or more generally cuts in the $x$-$y$ plane, the corresponding bias factor from haloes of fixed $\Mh$ or \Vpeak\ can be expressed as
\begin{equation}
    b_G=\iint_G dx dy\, p(x,y)b(x,y),
\end{equation}
where $G$ indicates the region defined by the cuts.

To proceed, we make the following assumptions -- (1) The joint distribution of galaxy and halo properties follows a 2-dimensional (2D) Gaussian $p(x,y)$, characterised by the centre $(x_c,y_c)$, standard deviations $\sigma_x$ and $\sigma_y$, and the correlation $\rho$ between $x$ and $y$. We can take $(x_c,y_c)=(0,0)$ by shifting $x$ and $y$. A 2D Gaussian function is a reasonable approximation for the distributions seen in Fig.\ref{fig:gh_m} and Fig.~\ref{fig:gh_vp}, which also follows the Taylor expansion of the logarithmic of the distribution function to the second order. (2) Galaxy property $y$ has a strong dependence on the halo assembly property $x$, and only weakly on other halo assembly variables. That is, any galaxy assembly bias effect is a result of inheriting from the assembly bias effect in halo property $x$, not other halo assembly properties. With this assumption, $b(x,y)$ has no additional dependence other than $x$, simplifying  to $b(x,y)=b(x)$. (3) At the fixed $\Mh$ (or $\Vp$), halo  bias factor is linear with respect to halo property $x$, 
$b(x)=kx+b_c$, the first order approximation from Taylor expansion. The value $b_c$ is the bias factor at $x=0$, which is also the average halo bias factor for haloes at mass $\Mh$ (or $\Vp$). The slope $k$ is the first derivative of $b$ with respect to $x$, $k=\partial b/\partial x$. In Appendix~\ref{sec:appendixb}, we present and test an extension of the model by loosing the assumptions (2) and (3),
and the tests confirm that the simplification here is reasonable.

For the top $f$ fraction of haloes with the highest $x$ and that of galaxies with the highest $y$, the distribution of halo property $x$ is shown in the bottom panel of Fig.~\ref{fig:demo} as the dashed curve and red curve, respectively. They are the projections of the region inside the dashed curve and that in red. Clearly the selected halo and galaxy samples differ in the mean halo property $x$, lower for the galaxy sample. The mean values can be calculated as
\begin{equation}
\label{eqn:avex}
\langle x\rangle_{x>t\sigma_x}  = 
\frac{\int_{t\sigma_x}^{+\infty}dx \int_{-\infty}^{+\infty} dy\, x p(x,y)
}{\int_{t\sigma_x}^{+\infty}dx \int_{-\infty}^{+\infty} dy\, p(x,y)} 
 = 
\frac{\sigma_x \exp(-t^2/2)}{\int_t^{+\infty} dv \exp(-v^2/2)}
\end{equation}
and
\begin{equation}
\label{eqn:avey}
\langle x\rangle_{y>t\sigma_y}  = 
\frac{\int_{t\sigma_y}^{+\infty}dy \int_{-\infty}^{+\infty} dx\, x p(x,y)
}{\int_{t\sigma_y}^{+\infty}dy \int_{-\infty}^{+\infty} dx\, p(x,y)} 
 = 
\frac{\rho\sigma_x \exp(-t^2/2)}{\int_t^{+\infty} dv \exp(-v^2/2)},
\end{equation}
where $t$ is determined by having the correct fraction $f$,
\begin{equation}
\label{eqn:ft}
 f=\int_t^{+\infty} dv \frac{1}{\sqrt{2\pi}}\exp(-v^2/2).
\end{equation}
The bias factors for the halo and galaxy samples are then
\begin{equation}
\label{eqn:bh}
b_h = 
\langle b\rangle_{x>t\sigma_x} = k \langle x\rangle_{x>t\sigma_x} + b_c
\end{equation}
and
\begin{equation}
\label{eqn:bg}
b_g = 
\langle b\rangle_{y>t\sigma_y} = k \langle x\rangle_{y>t\sigma_y} + b_c.
\end{equation}

We can characterise the assembly bias effect by the fractional difference between the bias factor of the selected halo/galaxy sample and the average halo bias factor at $\Mh$ (or $\Vp$), $\delta_h^b=(b_h-b_c)/b_c$ and $\delta_g^b=(b_g-b_c)/b_c$. Based on equations~(\ref{eqn:avex})--(\ref{eqn:bg}), we have
\begin{equation}
\label{eqn:deltab}
\delta_g^b=\rho\delta_h^b.
\end{equation}
That is, the assembly bias effect of the galaxy sample is weaker than that of the halo sample, by a factor equal to the correlation coefficient of the galaxy and halo property. Only in the case that galaxy and halo properties are tightly correlated (with zero scatter; $|\rho|=1$) does halo assembly bias effect completely translate to galaxy assembly bias effect. The connection in equation~(\ref{eqn:deltab}) is also valid for samples defined by the property range bounded by two percentiles (i.e. bin samples rather than threshold samples considered here).

To test how well the simple model works, we choose a pair of halo and galaxy properties to construct the halo and galaxy samples. We then measure the two-point correlation functions (2PCFs) of the halo and galaxy samples in each $\Mh$ and $\Vp$ bin. To reduce the uncertainty, the large scale bias factor of a given halo sample is derived from the ratio of the halo-matter two-point cross-correlation function and the matter auto-correlation function \citep[e.g.][]{Xu18}, $b_h=\xi_{hm}(r)/\xi_{mm}(r)$, averaged over scales of 5-18 $\hinvMpc$. The bias factor for the galaxy sample is similarly derived. We consider samples based on halo formation time \ahalf\ and galaxy colour/SFR/sSFR, which show $\Mh$ ($\Vp$) dependent correlation coefficient (Fig.~\ref{fig:gh_pearson}). The results are shown in Fig.~\ref{fig:gabhab}. 
In the top-left panel, the thick dotted red (blue) curves show the assembly bias quantity $\delta_h^b$ for the 50 per cent oldest (youngest) haloes as a function of $\Mh$. The solid red (blue) curves are $\delta_g^b$ for the 50 per cent reddest (bluest) central galaxies. Both quantities decrease with increasing halo mass, i.e. the assembly bias effect becomes weaker for more massive haloes. The thin dashed curves are the same as the dotted curves but modulated by the correlation coefficient between colour and \ahalf, i.e. $\rho\delta_h^b$, the prediction for $\delta_g^b$ from the simple model. The model works well in reproducing the halo mass dependent galaxy assembly bias effect based on halo assembly bias and the galaxy-halo correlation. The middle-left and bottom-left panels are for galaxies selected according to SFR and sSFR. The right panels show the assembly bias effect as a function of $\Vp$. In all the cases, $\delta_g^b$ can be well described by $\rho \delta_h^b$, which supports the effectiveness of the simple model.

The success of the model suggests that an easy recipe could be developed to incorporate galaxy assembly bias into the halo model. The contribution of central galaxies to the galaxy bias factor, in its full form in the simple model, is
\begin{equation}
b_g = b_c + \frac{\partial b}{\partial x} \left[x_c + \sigma_x \frac{\rho}{\sqrt{2\pi}} \exp(-t^2/2)/f\right],
\label{eqn:bg_simmod}
\end{equation}
where $t=t(f)$ is from equation~(\ref{eqn:ft}). As an example, let us use the halo mass $\Mh$ as the primary variable in the halo model and consider a galaxy property (colour) that correlates with halo formation time (\ahalf\ or $\log$\ahalf). In the halo model, besides the average halo bias $b_c$, we also need to know how the halo bias changes with \ahalf\ ($\partial b/\partial x$), the mean value of \ahalf\ ($x_c$), and the scatter in \ahalf\ ($\sigma_x$), all as a function of $\Mh$. As usual, we can construct fitting formulae for those four quantities based on $N$-body simulations. The quantities $f$ and $\rho$ belong to the description of the galaxy-halo relation, which can be parameterised. For $f$, it is simply the occupation fraction for haloes at $\Mh$. For $\rho$, the results in Fig.~\ref{fig:gh_pearson} suggest that a quadratic form with $\Mh$ would suffice. To compute the galaxy bias factor for a galaxy sample, we only need to perform a 1D integral over halo mass. Certainly it is straightforward to include assembly bias effect by populating dark matter haloes in $N$-body simulations. However, the above proposal has its virtue for analytic calculations in theoretical investigations.

The simple model can also be applied to observation to infer the correlation between galaxy and halo properties. \citet{Lin16} construct samples of central galaxies from the Sloan Digital Sky Survey data and study the assembly bias effect from the 2PCF measurements. Early and late galaxy samples are defined according to either star formation history (SFH) or sSFR. Weak lensing measurements are used to verify that the host haloes of the early and late galaxies are of similar halo mass (around $10^{12}\hinvMsun$). \citet{Lin16} compare the difference in the early and late galaxy clustering to that in the early and late formed haloes, and do not find evidence for galaxy assembly bias. For SFH-based galaxy samples, they find the ratio of the early to late galaxy bias factor to be $1.00\pm 0.12$ (see their fig.5). If we take the mean of the bias factors of the two galaxy samples as the average halo bias and the uncertainty comes from two similar error bars added in quadrature, the measurement gives $\delta_g^b=0$ with an uncertainty 0.085. For halos around $10^{12}\hinvMsun$, $\delta_h^b\sim 0.25$, with early- and late-forming haloes (Fig.~\ref{fig:gh_pearson}). 
 The coefficient of the correlation between SFH and halo formation time is then inferred to be $\rho=\delta_g^b/\delta_h^b=0.00\pm0.34$. For the sSFR-based samples, the ratio of the early to late galaxy bias factor is $1.07\pm 0.14$ (their fig.5). We infer $\delta_g^b=0.034\pm 0.099$, and with $\delta_h^b\sim 0.25$ the coefficient of the correlation between sSFR and halo formation time is constrained to be $\rho=0.14\pm 0.40$. For both cases, the correlation between galaxy and halo property is consistent with being small. It implies that galaxy SFH and sSFR at most only loosely track halo formation. Similar measurements with large samples can reduce the uncertainty in the inferred correlation, which would help test galaxy formation models (e.g. by comparing to those in Fig.~\ref{fig:gh_pearson}).

\section{Summary and Discussion}
\label{sec:summary}

Properties in galaxies residing in haloes of the same mass may have a dependence on certain aspects of halo assembly or formation history, which is termed as galaxy assembly bias. Studying galaxy assembly bias and its relation to halo properties can help improve the halo model of galaxy clustering and yield insights in galaxy formation and evolution. Using the Illustris cosmological hydrodynamic galaxy formation simulation, we investigate the central galaxy assembly bias effect through studying the relation among a set of galaxy and halo properties. 

The main results can be summarised as follows.
\begin{itemize}
    \item[(1)] Central galaxy stellar mass $\Mstar$ has a tighter relation with $\Vp$ than with $\Mh$, manifested by the smaller scatter in $\Mstar$ at fixed $\Vp$ than that at fixed $\Mh$. Once the assembly effect of $\Mstar$ on $\Vp$ is included, $\Mstar$ shows nearly no correlation with any other halo assembly properties. 
    
    \item[(2)] The correlations between halo assembly properties and other galaxy properties also appear cleaner if studied at fixed $\Vp$, which reveal that galaxy SFR, sSFR, and colour mainly correlate with halo formation time (and to a less extent with halo concentration).
    
    \item[(3)] A simple model is presented to show the relation between galaxy and halo assembly bias, which is linked by the correlation coefficient of the galaxy and halo property in consideration.
\end{itemize}

The Illustrius simulation produces a  relation between central galaxy stellar mass and halo mass ($\Mstar$--$\Mh$) similar to that inferred from observation. We find that the scatter in the relation is closely related to halo assembly properties. For example, at fixed $\Mh$, haloes of higher $\Vp$ or earlier formation tend to host galaxies of higher $\Mstar$. If we choose $\Vp$ to be the primary halo variable, the scatter in $\Mstar$ at fixed $\Vp$ is reduced compared to that at fixed $\Mh$. Remarkably, once switched to $\Vp$, $\Mstar$ appears to have nearly no dependence on other halo assembly properties, at least for those considered in our study (including halo concentration, formation time, spin, accretion rate, and specific accretion rate). The property $\Vp$, which is an indication of the maximum potential depth over the assembly history of haloes, is able to capture almost all the assembly effect in galaxy stellar mass. The results are in broad agreement with the study using the EAGLE simulation in terms of the $z=0$ maximum halo circular velocity \citep{Matthee17}. The reason for $\Vp$ to be the fundamental property in determining $\Mstar$ is likely related to the accretion and response of baryons in the gravitational potential. As for the scatter, the simulation noise in Illustris does not contribute much \citep{Genel18}. It could be related to chaotic or stochastic processes in star formation and feedback \citep{Matthee17,Genel18}. Further study is needed to understand the cause of the correlation between $\Mstar$ and $\Vp$ and the origin of the scatter.

We present the correlation between each pair of galaxy and halo property in terms of the Pearson correlation coefficient. Besides $\Mstar$, the other galaxy properties (SFR, sSFR, and colour) show a more consistent and clear trend with $\Vp$ than with $\Mh$. At fixed $\Vp$, those other galaxy properties related to star formation are found to mainly correlate with halo formation time and concentration, with stronger correlation with the former. The relatively nice behaviour in the correlations with galaxy properties in the $\Vp$-based investigation suggests that it would be advantageous to use {\it $\Vp$ as the primary variable} in halo model of galaxy clustering, in particular in modelling stellar-mass-based samples 
(while the combination of or interpolation between halo mass and circular velocity quantities remains as a possibility to investigate, in the spirit of \citealt{Lehmann17}). To further model SFR-, sSFR-, or colour-selected samples of galaxies, our investigation suggests to introduce {\it halo formation time as the secondary halo variable} (and to a less extent, halo concentration). This is in line with the age-matching model of \citet{Hearin13}, who assumes monotonic mapping between galaxy colour and some variant of halo formation time at fixed galaxy luminosity (or stellar mass). That is, there exists a perfect correlation between the galaxy and halo property. Our investigation shows, however, that the correlation coefficient between galaxy SFR/sSFR/colour and halo formation time should be included as one important ingredient in the model.

The Illustris simulation is able to reproduce the observed star-forming main sequence reasonably well \citep{Sparre15}. Here we find that at fixed $\Mh$ or $\Vp$ the relation between SFR and $\Mstar$ of central galaxies follows the star-forming main sequence. Interestingly, compared with the SFR--$\Mstar$ relation, SFR shows a tighter correlation with sSFR ($\rho\sim 0.9$) over the full halo $\Mh$ or $\Vp$ range. It is necessary to test its validity with observations and study its origin by tracking SFR and stellar mass growth of individual galaxies in simulations. The correlation between SFR and sSFR or between any pair of galaxy properties ($\Mstar$, SFR, sSFR, and colour) considered here cannot be explained solely by their common dependence on one halo assembly variable. Baryonic processes in galaxy formation (like star formation and feedback) likely play a major role in shaping such correlations.

For the effect of assembly bias on galaxy clustering, at fixed $\Mh$ or $\Vp$, we come up with a simple model to relate the bias factors of a galaxy sample and the corresponding halo sample, which are connected by the correlation coefficient of the galaxy and halo properties used to define the two samples. It gives a reasonable description for the samples constructed with the simulation. It suggests a simple prescription to incorporate galaxy assembly bias into the halo model. By applying the simple model to the galaxy clustering measurements in \citet{Lin16}, we infer that the correlation between SFH/sSFR and halo formation time is consistent with being weak ($\rho\sim 0$--0.14). The simple model can be further tested with other hydrodynamic simulations, like EAGLE \citep{Schaye15} and IllustrisTNG \citep{Nelson18}, and semi-analytic models, which can also provide further insights on parameterising the correlations between galaxy and halo properties. While our study in this paper focuses on central galaxies, we plan to carry out similar investigations for satellite galaxies to complete the picture of galaxy assembly bias at the halo level to help improve the halo model.

\section*{Acknowledgements}
We thank David Weinberg, Jeremy Tinker and H\'elion du Mas des Bourboux for useful comments. The support and resources from the Center for High Performance Computing at the University of Utah are gratefully acknowledged. This research was supported by the Munich Institute for Astro- and Particle Physics (MIAPP) of the DFG cluster of excellence ``Origin and Structure of the Universe'', and ZZ thanks the hospitality of MIAPP, where the revision of the manuscript was undertaken.



\appendix
\section{The \texorpdfstring{$\Mstar$--\Vmax}\ Relation and the Baryon Effect on the \texorpdfstring{$\Mstar$--\Vpeak}\ Relation}
\label{sec:appendixa}

\begin{figure*}
	\centering
	\begin{subfigure}[h]{0.48\textwidth}
		\includegraphics[width=\textwidth]{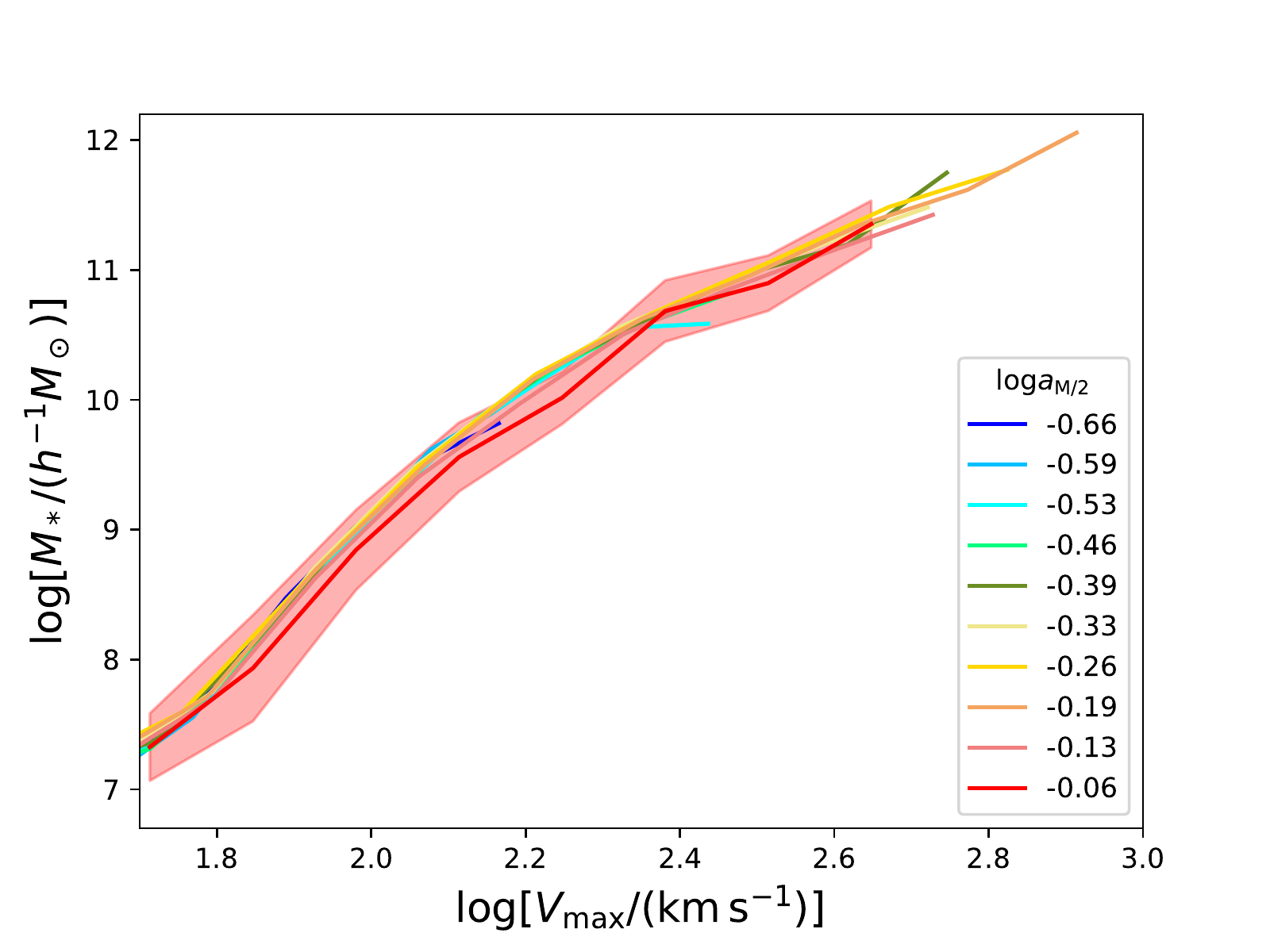}
	\end{subfigure}
	\hfill
	\begin{subfigure}[h]{0.48\textwidth}
        \includegraphics[width=\textwidth]{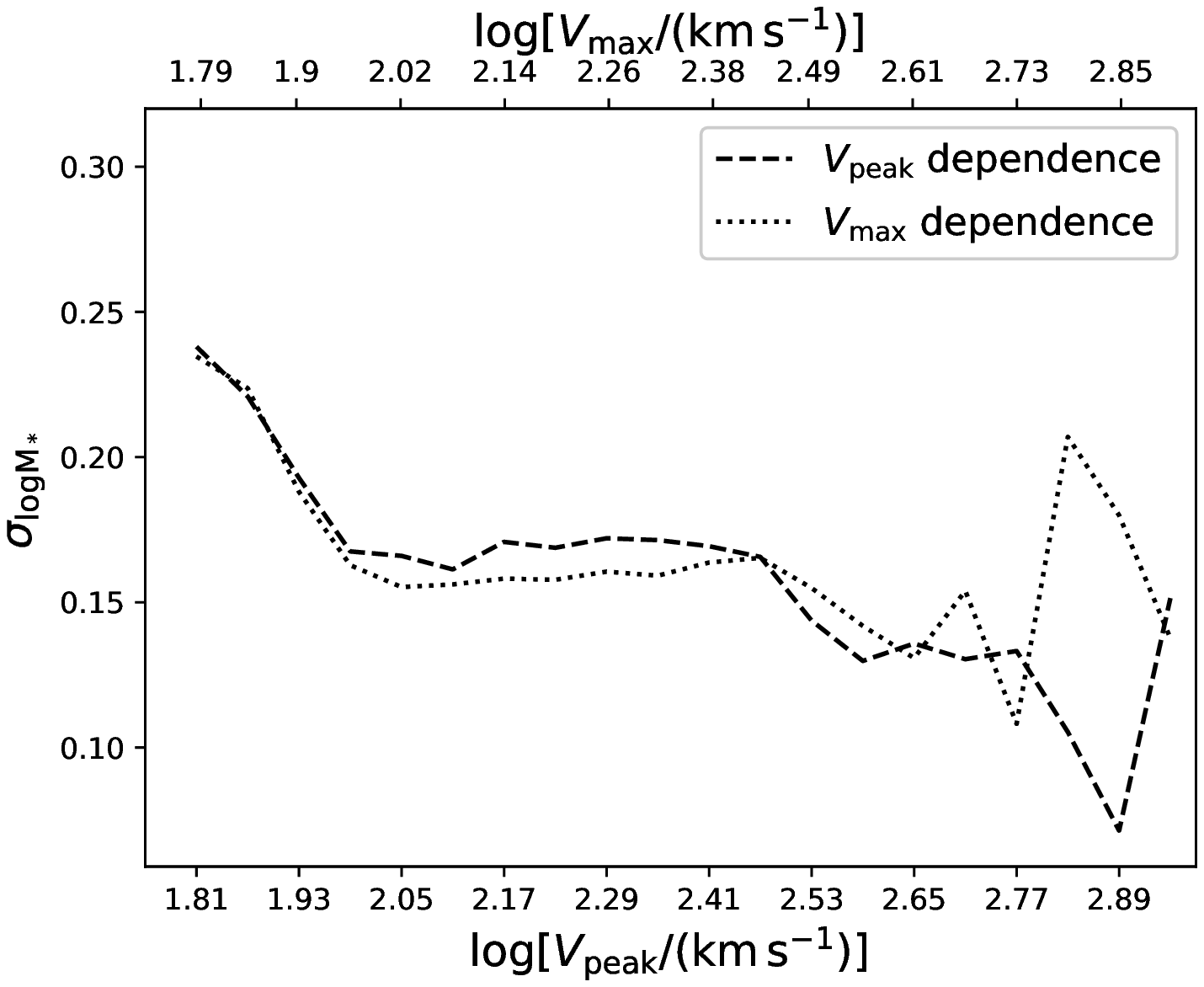}
	\end{subfigure}
	\hfill
\caption{
Left: Same as the bottom-right panel of Fig.~\ref{fig:mstar_plot}, but showing the dependence of $\Mstar$ on \Vmax\ instead of \Vpeak.
Right: Similar to Fig.~\ref{fig:mstarscatter}, but comparing the standard deviations in $\log\Mstar$ as a function of \Vpeak\ and \Vmax. The correspondence between \Vpeak\ and \Vmax\ comes from fitting a mean relation (see text for detail).
}
\label{fig:mstar_vmax}
\end{figure*}

\begin{figure*}
	\centering
	\begin{subfigure}[h]{0.48\textwidth}
		\includegraphics[width=\textwidth]{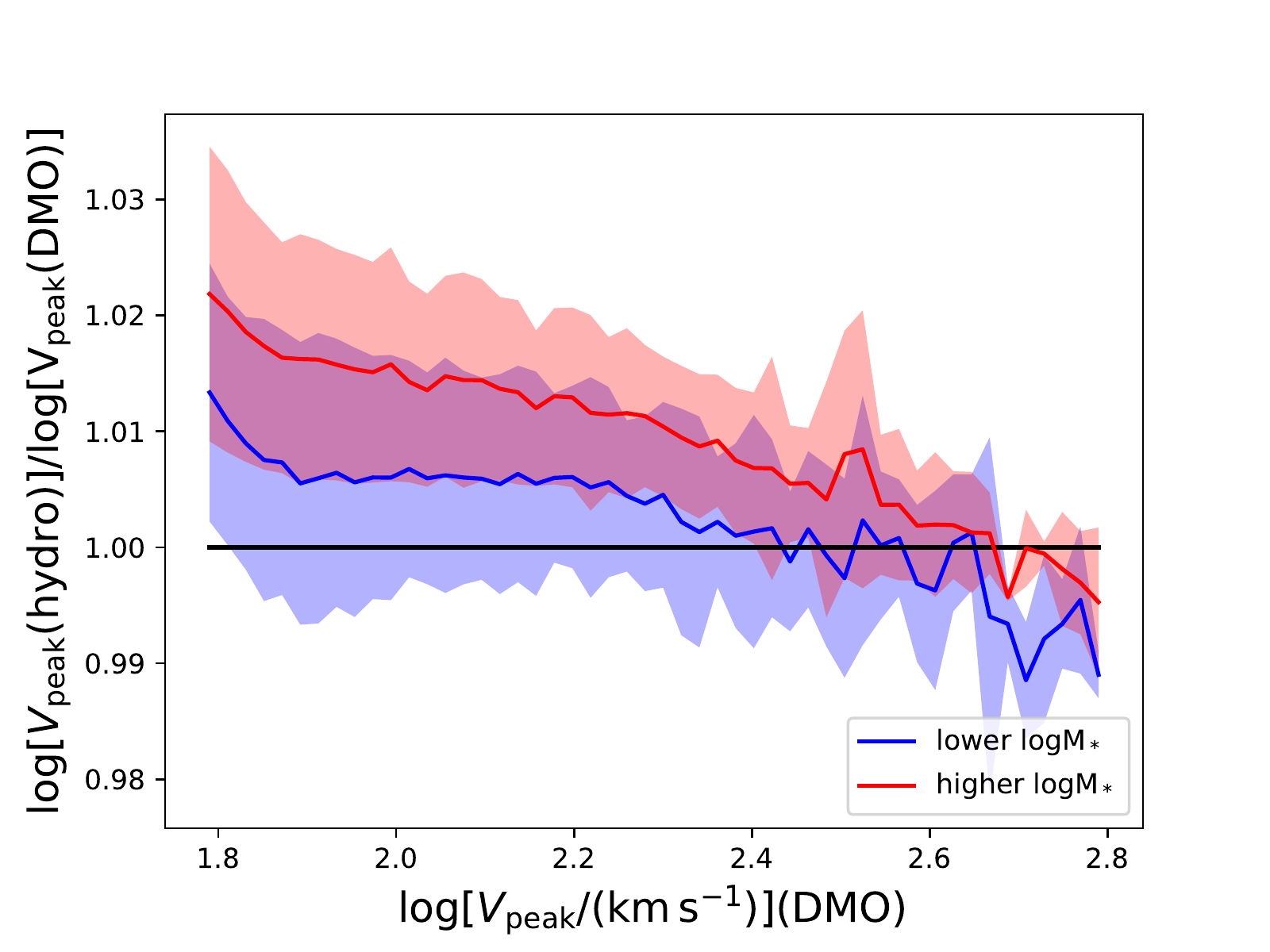}
	\end{subfigure}
	\hfill
	\begin{subfigure}[h]{0.48\textwidth}
        \includegraphics[width=\textwidth]{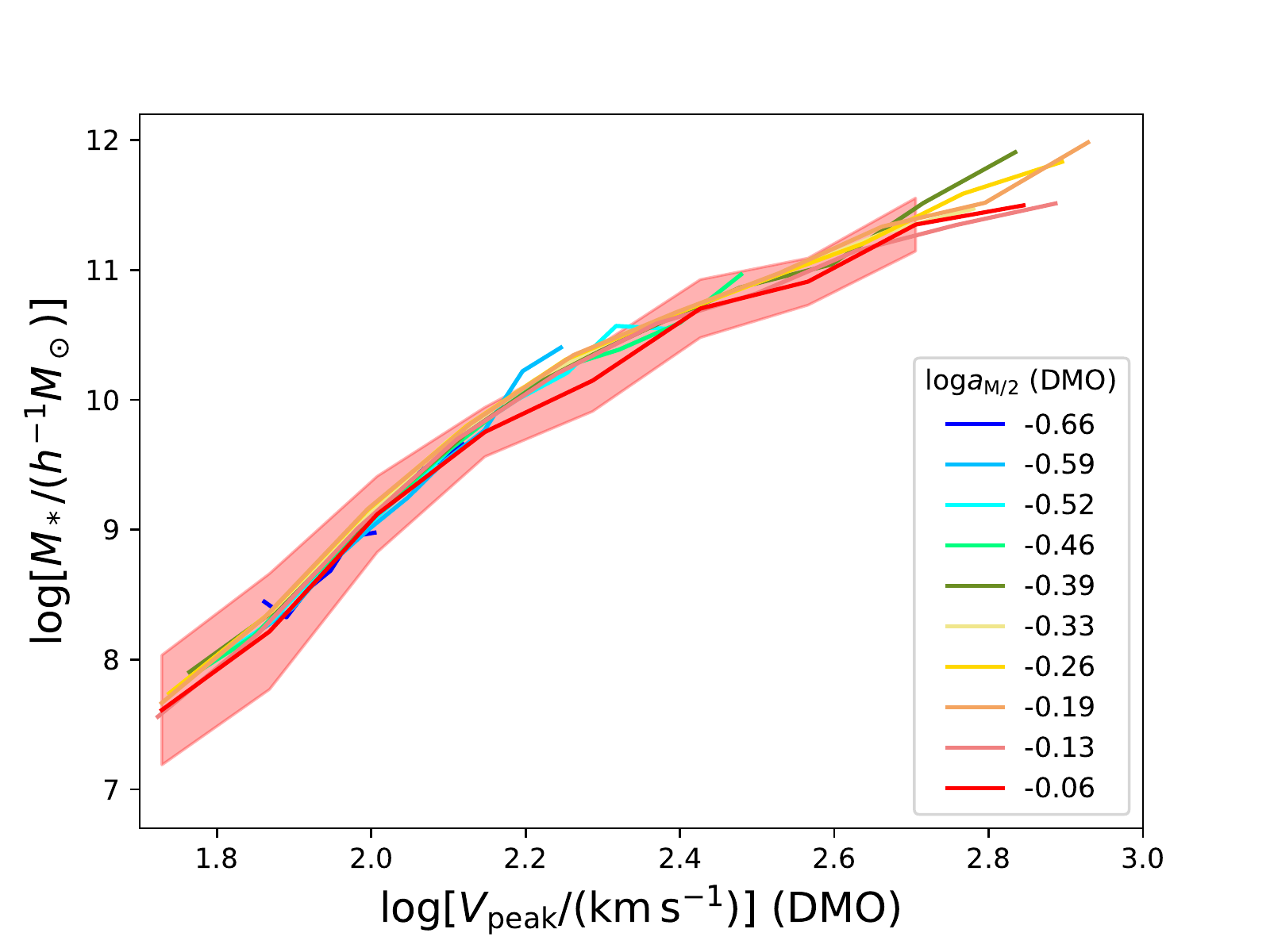}
	\end{subfigure}
	\hfill
\caption{Effect of baryonic physics on \Vpeak. 
Left: ratio of \Vpeak\ from the full physics (hydro) simulation to that from the DMO simulation as function of DMO \Vpeak. Red and blue solid are \Vpeak\ ratio for haloes with central stellar mass in the upper and lower halves at fixed DMO \Vpeak. Standard deviations are shown by red and blue bands. Note that hydro \Vpeak\ only slightly differ from DMO \Vpeak, at the per cent level. Right: similar to the bottom-right panel of Fig.~\ref{fig:mstar_plot}, but with \Vpeak\ and \ahalf\ from the DMO simulation.
\label{fig:DMO}
}
\end{figure*}

In section~\ref{sec:mstardep}, we show that central galaxy $\Mstar$ has a tight correlation with \Vpeak\ and there is no further dependence on  \ahalf\ at fixed \Vpeak. The other velocity quantity commonly considered is \Vmax, the present-day maximum circular velocity of a halo. While both \Vpeak\ and \Vmax\ characterise the depth of the potential well of a halo, the computation of \Vmax\ in a simulation is more straightforward, without tracking the whole formation history of a halo. We find that the $\Mstar$--\Vmax\ relation is similar to the $\Mstar$--\Vpeak\ relation, as shown in the left panel of Fig.~\ref{fig:mstar_vmax}. However, the $\Mstar$--\Vmax\ relation has a residual dependence on \ahalf, with on average higher stellar mass in older haloes. This suggests that \Vpeak\ better captures the effect of halo formation time on the final central galaxy stellar mass than \Vmax. 

In the right panel of Fig.~\ref{fig:mstar_vmax}, we show the standard deviation in $\log\Mstar$ as a function of \Vpeak\ and \Vmax, respectively. We fit a straight line to derive the mean relation between $\log$\Vpeak\ and $\log$\Vmax, and the correspondence between \Vpeak\ and \Vmax\ in the figure comes from such a fit. The \Vmax\ dependence of the scatter in $\Mstar$ only slightly differs from the \Vpeak\ dependence. In particular, compared to the \Vpeak\ dependence of the scatter in $\Mstar$, that for the \Vmax\ dependence is smaller (higher) for haloes with \Vpeak\ below (above) $\sim$300 ${\rm km\, s^{-1}}$. Although \Vpeak\ explains the dependence of $\Mstar$ on formation time better than \Vmax, the tighter $\Mstar$--\Vmax\ relation in lower \Vpeak\ haloes indicates that $\Mstar$ could also depend on variables other than halo formation time (e.g. halo environment).

In section~\ref{sec:mstardep}, the quantity \Vpeak\ used in the analyses is from the full physics simulation of Illustris. One may wonder whether baryons are partly responsible for the tight $\Mstar$--\Vpeak\ relation, as gas cooling with subsequent star formation can deepen the potential and thus change \Vpeak. We explore the baryon effect on the $\Mstar$--\Vpeak\ relation by replacing \Vpeak\ and \ahalf\ with those in the dark-matter-only (DMO) simulation, after matching Rockstar haloes found in the Illustris-2-Dark simulation to those in the Illustris-2 simulation. In the left panel of Fig.~\ref{fig:DMO}, we show the ratio of full physics (hydro) \Vpeak\ to the DMO \Vpeak\ as a function of DMO \Vpeak. At each DMO \Vpeak\ bin, we further divide haloes into two halves, with high and low $\Mstar$, respectively. Haloes with more massive central galaxies at fixed DMO \Vpeak\ tend to have higher hydro-to-DMO \Vpeak\ ratio, reflecting the effect of baryons on deepening the potential. The difference between hydro \Vpeak\ and DMO \Vpeak\ is larger in haloes of lower DMO \Vpeak. However, the overall difference is small, e.g. below $\sim$3 per cent. This difference plays a negligible role in the $\Mstar$--\Vpeak\ relation, as shown by comparing the right panel of Fig.~\ref{fig:DMO} to the bottom-right panel of Fig.~\ref{fig:mstar_plot}.

\section{A More General Form of Central Galaxy Assembly Bias}
\label{sec:appendixb}

\begin{figure*}
	\centering
	\begin{subfigure}[h]{0.48\textwidth}
		\includegraphics[width=\textwidth]{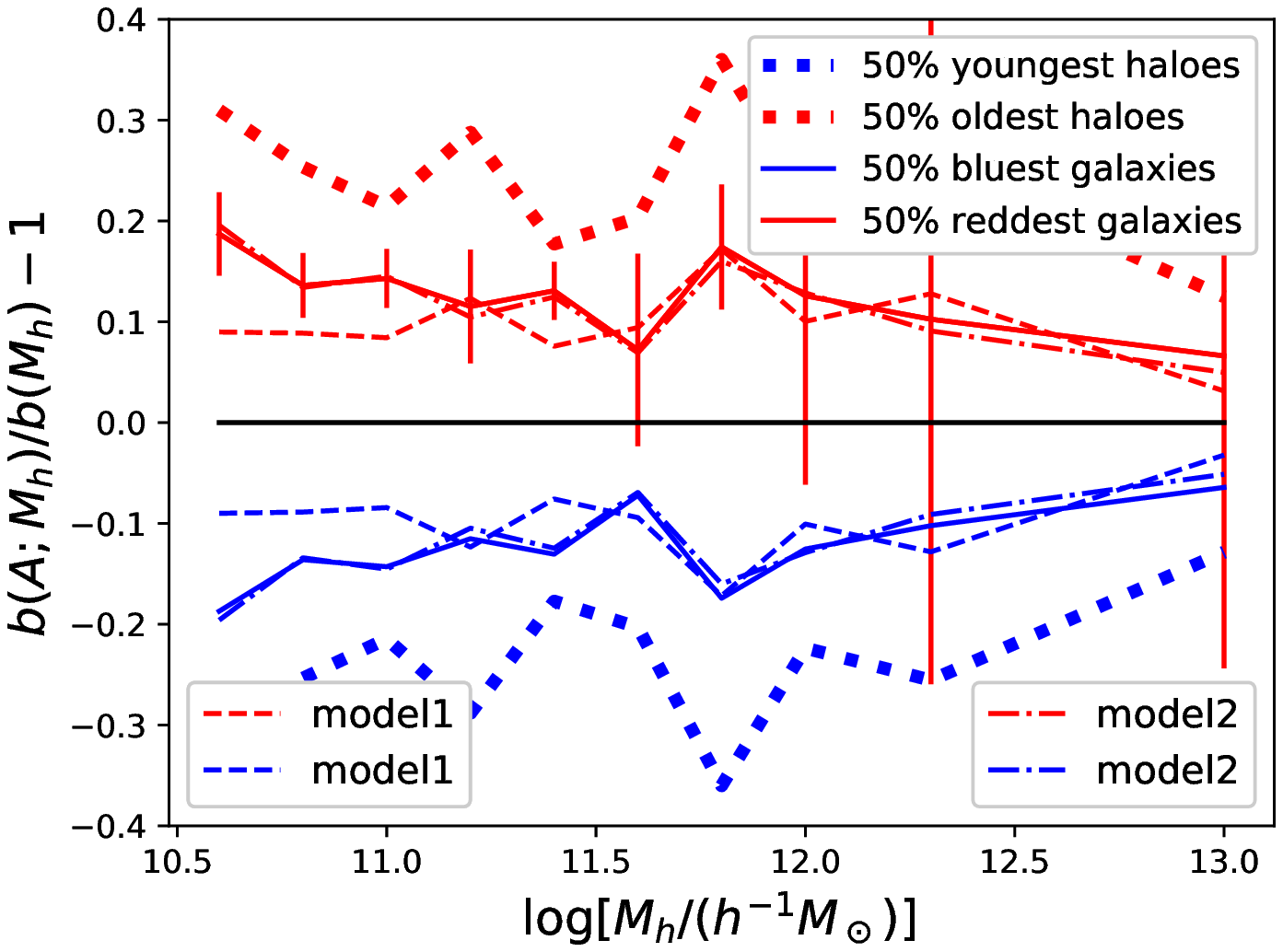}
	\end{subfigure}
	\hfill
	\begin{subfigure}[h]{0.48\textwidth}
                \includegraphics[width=\textwidth]{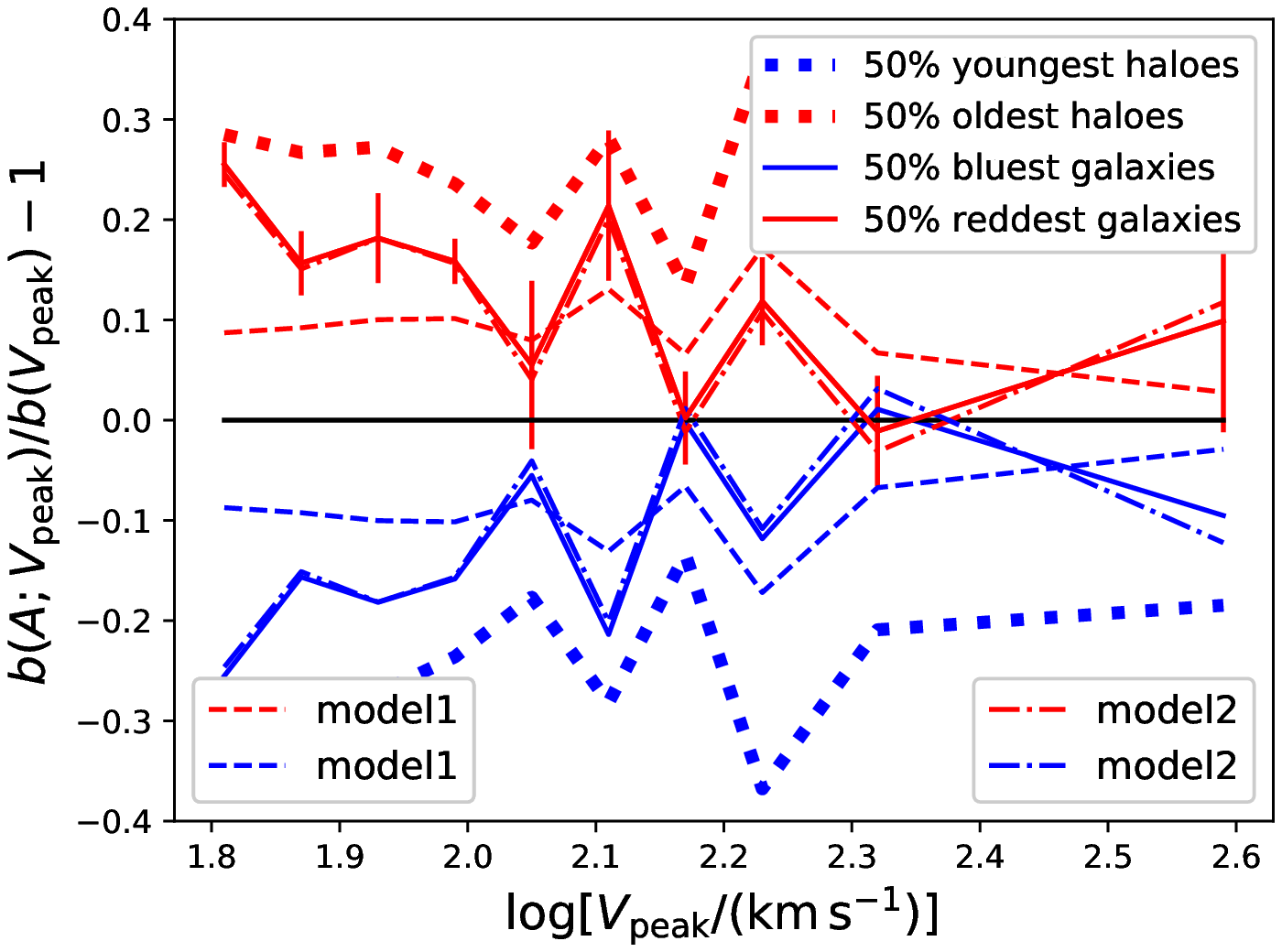}
	\end{subfigure}
	\hfill
	\begin{subfigure}[h]{0.48\textwidth}
                \includegraphics[width=\textwidth]{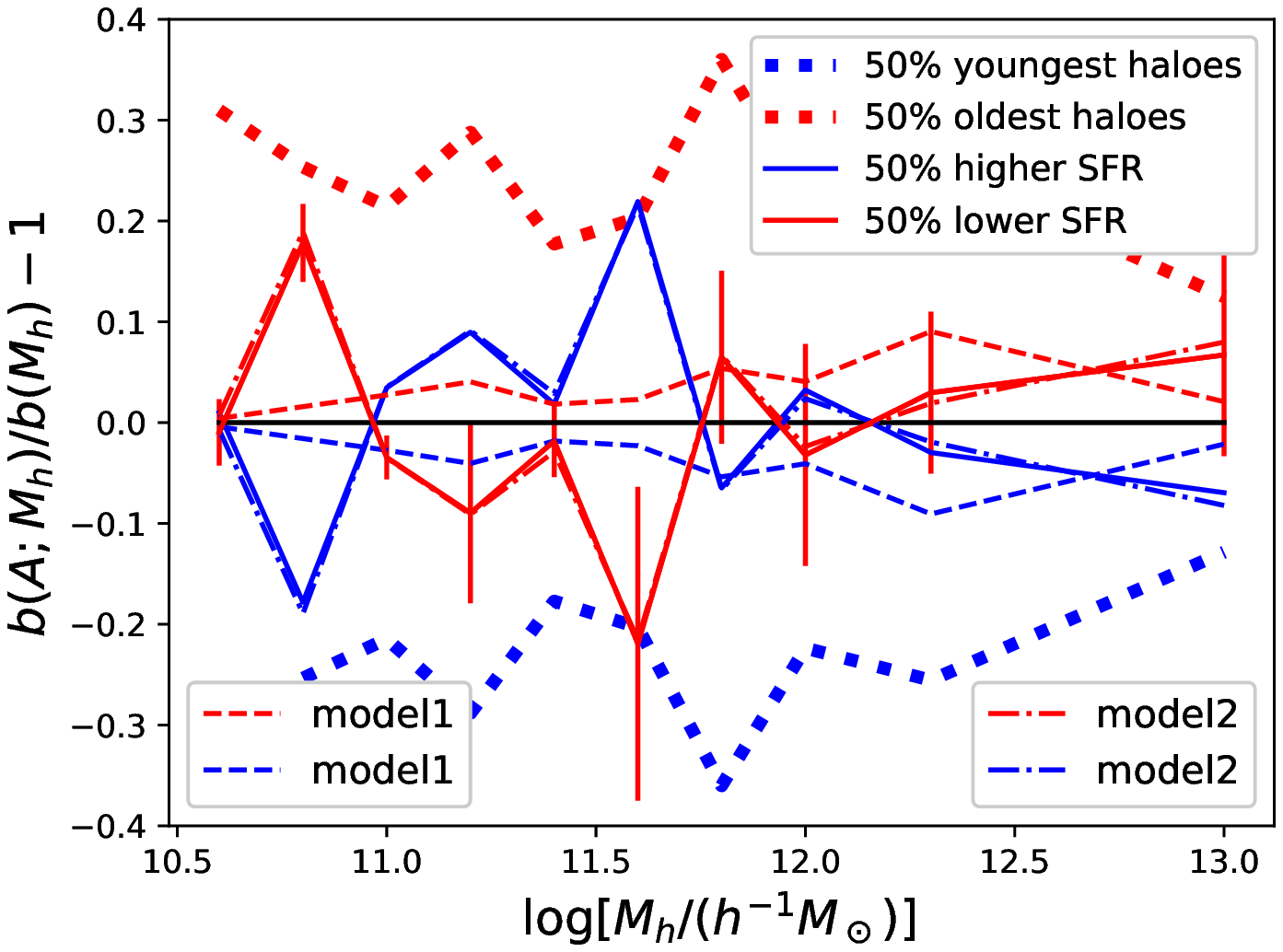}
	\end{subfigure}
	\hfill
	\begin{subfigure}[h]{0.48\textwidth}
		\includegraphics[width=\textwidth]{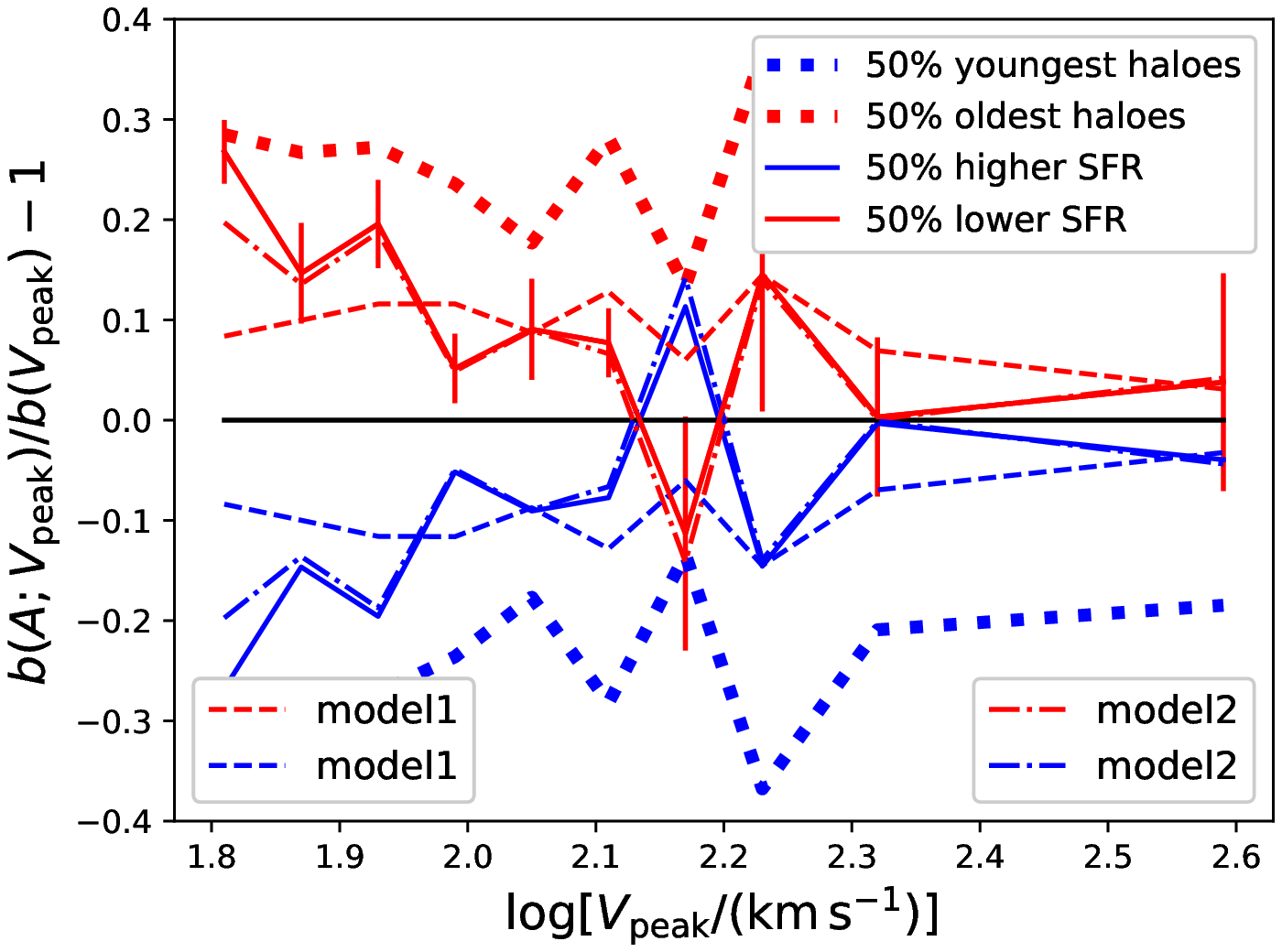}
	\end{subfigure}
	\hfill
	\begin{subfigure}[h]{0.48\textwidth}
                \includegraphics[width=\textwidth]{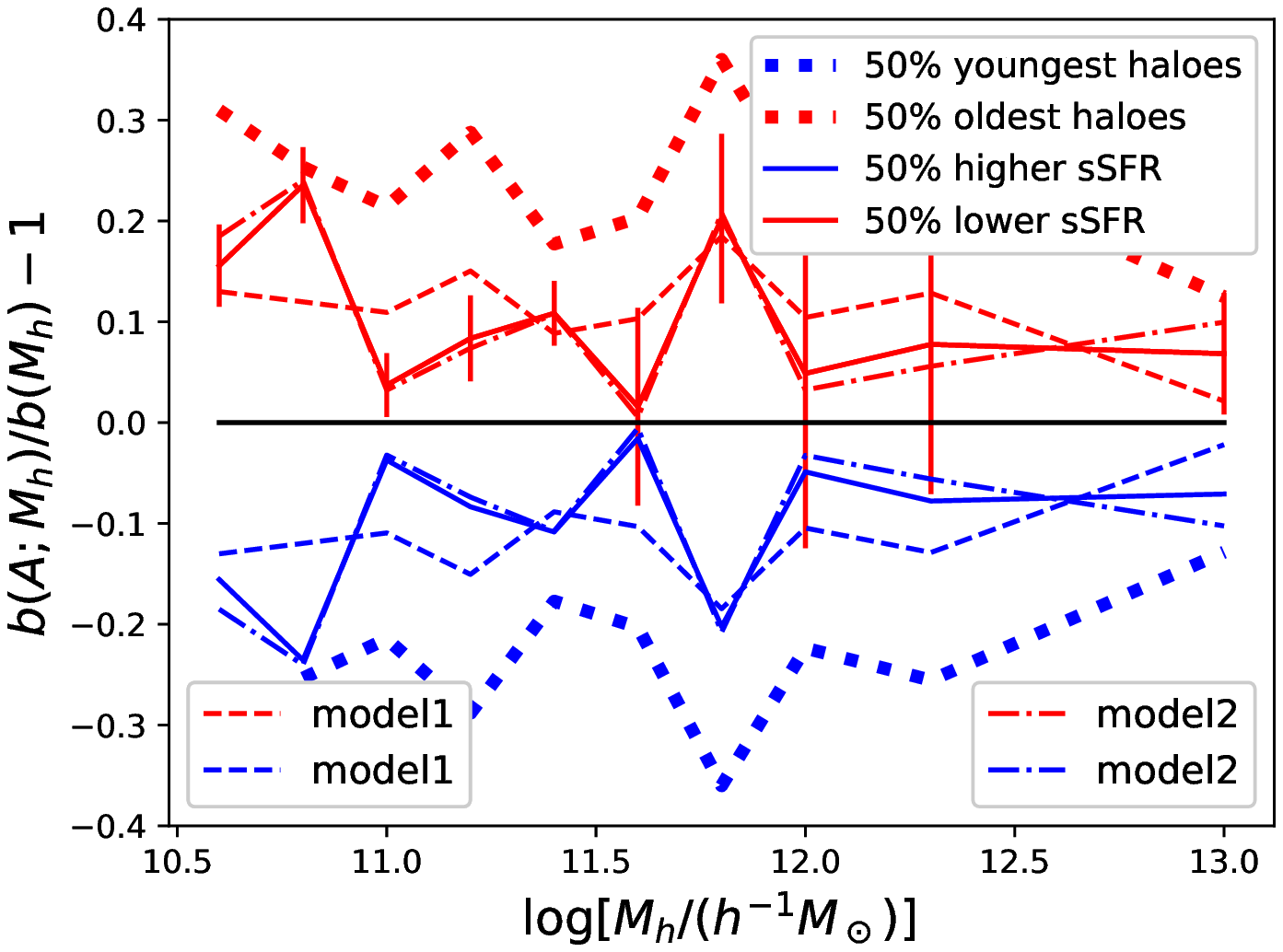}
	\end{subfigure}
	\hfill
	\begin{subfigure}[h]{0.48\textwidth}
                \includegraphics[width=\textwidth]{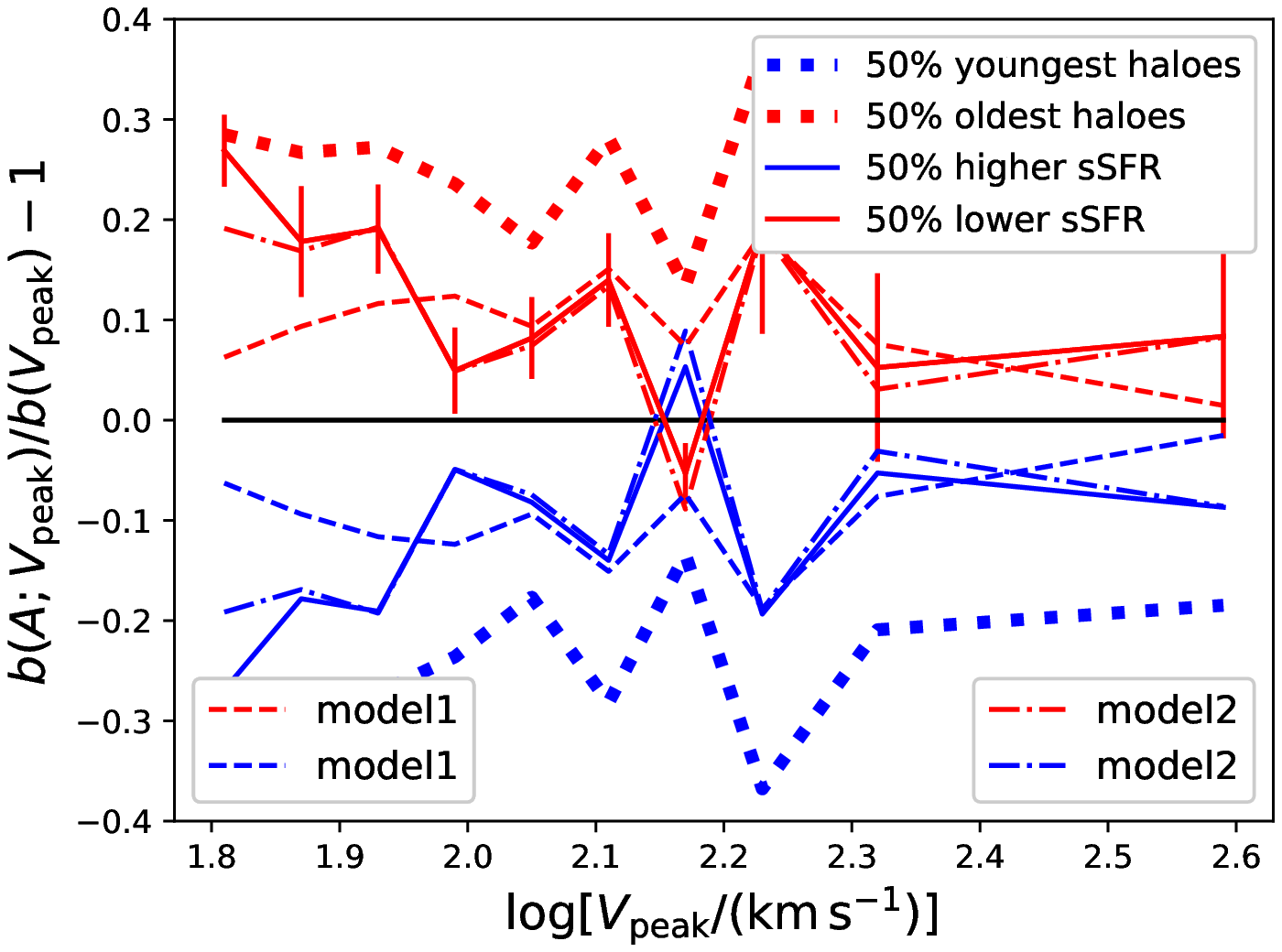}
	\end{subfigure}
	\hfill
\caption{
Same as Fig.~\ref{fig:gabhab}, but in comparison to the predictions from the simple model (dashed, labelled as `model 1') in section~\ref{sec:gab}, those from a more general model (dash-dotted, labelled as `model 2') are added. See text for detail.
}
\label{fig:gabmodel2}
\end{figure*}

In section~\ref{sec:gab} we develop a simple model to describe central galaxy assembly bias based on halo assembly bias and galaxy-halo relation. In that model, bias is assumed to depend only on one halo property at fixed halo mass. The model works reasonably well based on the tests with Illustris simulation. Since halo assembly bias is found to have a multi-variate dependence \citep[e.g.][]{Mao18,Xu18,Han19}, it is useful to extend the simple model for more general applications. Instead of increasing the dimensionality of the model and describing halo assembly bias as a function of multiple halo assembly variables at fixed halo mass, we choose to model halo bias as a function of one halo assembly property ($x$) and one galaxy property ($y$). That is, the bias dependence on any additional halo assembly properties is absorbed into that on the galaxy property (as galaxy property can be related to those halo properties). Additionally, the bias factor is assumed to be linearly dependent on halo and galaxy property, $b(x,y)=k_x x + k_y y + b_c$. Therefore, the contours of constant bias now have the possibility of deviating from vertical in the top panel of Fig.~\ref{fig:demo}. The coefficient $k_y$ captures the dependence of $y$ on other halo properties and the scatter in the $y$-$x$ relation.

With this model, equations~(\ref{eqn:bh}) and (\ref{eqn:bg}) become
\begin{equation}
\label{eqn:bh_xy}
b_h = 
\langle b\rangle_{x>t\sigma_x} = k_x \langle x\rangle_{x>t\sigma_x} + k_y \langle y\rangle_{x>t\sigma_x} + b_c
\end{equation}
and
\begin{equation}
\label{eqn:bg_xy}
b_g = 
\langle b\rangle_{y>t\sigma_y} = k_x \langle x\rangle_{y>t\sigma_y} + k_y \langle y\rangle_{y>t\sigma_y} + b_c,
\end{equation}
where the new terms $\langle y\rangle_{x>t\sigma_x}$ and $\langle y\rangle_{y>t\sigma_y}$ can be calculated as
\begin{equation}
\label{eqn:avey_tx}
\langle y\rangle_{x>t\sigma_x}  = 
\frac{\int_{t\sigma_x}^{+\infty}dx \int_{-\infty}^{+\infty} dy\, y p(x,y)
}{\int_{t\sigma_x}^{+\infty}dx \int_{-\infty}^{+\infty} dy\, p(x,y)} 
 = 
\frac{\rho\sigma_y \exp(-t^2/2)}{\int_t^{+\infty} dv \exp(-v^2/2)}
\end{equation}
and
\begin{equation}
\label{eqn:avey_ty}
\langle y\rangle_{y>t\sigma_y}  = 
\frac{\int_{t\sigma_y}^{+\infty}dy \int_{-\infty}^{+\infty} dx\, y p(x,y)
}{\int_{t\sigma_y}^{+\infty}dy \int_{-\infty}^{+\infty} dx\, p(x,y)} 
 = 
\frac{\sigma_y \exp(-t^2/2)}{\int_t^{+\infty} dv \exp(-v^2/2)}.
\end{equation}

Similar to what we do in section~\ref{sec:gab} and based on equations~(\ref{eqn:avey_tx}), (\ref{eqn:avey_ty}), (\ref{eqn:bh_xy}), (\ref{eqn:bg_xy}), (\ref{eqn:avex}), and (\ref{eqn:avey}), we obtain the following relation between the fractional halo and galaxy assembly bias  ($\delta_h^b$ and $\delta_g^b$), 
\begin{equation}
\label{eqn:deltab_xy}
\delta_g^b=\frac{\rho+\gamma}{1+\rho\gamma}\delta_h^b,
\end{equation}
where $\gamma=(k_y\sigma_y)/(k_x\sigma_x)$ is an indicator of the relative importance of the bias dependence on galaxy and halo property. For $\gamma$=0 (e.g. with $k_y=0$), this model reduces to the simple one in section~\ref{sec:gab}, with the bias gradient along the $x$ direction (bias contours being vertical). In general, $|\gamma|< |\rho|$ corresponds to bias contours close to vertical, while $|\gamma|> |\rho|$ close to horizontal. In the following we will inverstigate which case represents that in the simulation by calculating $\gamma$.

We examine the performance of equation~(\ref{eqn:deltab_xy}) using the Illustris simulation and test the importance of $\gamma$. Same as in section~\ref{sec:gab}, halo assembly property ($x$) we consider is the formation time \ahalf, and galaxy properties ($y$) include colour, SFR, and sSFR. At a fixed $\Mh$ or \Vpeak\ bin, haloes are divided into four sub-samples, depending which lower/upper halves of $x$ and lower/upper halves of $y$ they belong to. We measure the bias factors of the sub-samples and derive $k_x$ and $k_y$ from fitting them with $b=k_x \langle x\rangle + k_y\langle y\rangle +b_c$. Together with the measured standard deviations $\sigma_x$ and $\sigma_y$, we infer $\gamma$ in equation~(\ref{eqn:deltab_xy}.) 

The results are shown in Fig.~\ref{fig:gabmodel2}. It is similar to Fig.~\ref{fig:gabhab}, except that a set of dash-dotted curves are added to show the predicted $\delta_g^b$ by equation~(\ref{eqn:deltab_xy}) (and that larger bins are used for haloes of high $\Mh$/\Vpeak\ to reduce the noise in the bias measurements for sub-samples). Clearly the more general model of equation~(\ref{eqn:deltab_xy}) leads to better agreements with the measured galaxy assembly bias, and in almost of the cases the predictions just fall on top of the measurements. In general, however, we find that $\gamma$ only plays a minor role in determining the fractional assembly bias, compared to $\rho$. For example, we find $\gamma=-0.13\pm 0.22$ with \ahalf\ and galaxy colour across all $\Mh$ bins, lower in amplitude than $\rho$ (about 0.3--0.5, as shown in Fig.~\ref{fig:gh_pearson}). Similar results are also obtained with other galaxy properties {\it that we used in Fig.~\ref{fig:gabhab}}. That is, as long as the halo assembly property we identify is closely connected to galaxy property, any additional dependence of bias (on other halo properties, through that on galaxy property) is weak. The result indicates that considering only the bias dependence of such a halo property is sufficient to describe the most useful part of halo bias in modelling galaxy bias and the simple model in section~\ref{sec:gab} would work reasonably well. It would be interesting to see how general this holds by testing with more galaxy and halo properties and by using different galaxy formation simulations.

If one would like to incorporate any additional bias dependence into the model, the counterpart of equation~(\ref{eqn:bg_simmod}) for the contribution of central galaxies to the galaxy bias factor reads
\begin{equation}
b_g = b_c + \frac{\partial b}{\partial x} \left[x_c + \Delta x_c + \sigma_x \frac{\rho_{\rm eff}}{\sqrt{2\pi}} \exp(-t^2/2)/f\right],
\end{equation}
where $\Delta x_c=\gamma(\sigma_x/\sigma_y)y_c$ and $\rho_{\rm eff}=\rho+\gamma$. Instead of parameterising $\rho$ (as a function of $\Mh$), we now only need to describe two quantities, the offset $\Delta x_c$ in halo assembly property and the effective correlation coefficient $\rho_{\rm eff}$. As with equation~(\ref{eqn:bg_simmod}), the expression follows the spirit of halo model -- $b_c$, $\partial b/\partial x$, $x_c$, and $\sigma_x$ are halo properties, while $\Delta x_c$, $\rho_{\rm eff}$, $t$, and $f$ describe the galaxy-halo relation, which includes the correlation and scatter between the galaxy and halo property and the galaxy occupation function.

To achieve more precise modelling, it is also possible to express halo bias as linear functions of more than two halo properties, instead of absorbing that into the dependence of galaxy property, and to include quadratic dependence of each halo property. While such extensions are straightforward in our framework, it requires multi-variate descriptions of halo and galaxy properties with a substantially increased number of parameters, making the model less tractable and less attractive. In practice, the simple model presented in section~\ref{sec:gab} would be a useful first step.


\bsp	
\label{lastpage}
\end{document}